\pdfoutput=1

\documentclass[12pt]{article}

\usepackage{amsfonts}
\usepackage{amsmath}
\usepackage{amssymb}
\usepackage{bigints}
\usepackage{booktabs}
\usepackage[nosort]{cite}
\usepackage{color}
\usepackage{dsfont}
\usepackage{float}
\usepackage{framed}
\usepackage{graphicx}
\usepackage{indentfirst}
\usepackage{mathrsfs}
\usepackage{multirow}
\usepackage{pdflscape}
\usepackage{setspace}
\usepackage{subdepth}
\usepackage{subfig}
\usepackage{titlesec}
\usepackage[dotinlabels]{titletoc}
\usepackage{wrapfig}
\usepackage[all]{xy}
\usepackage{young}
\usepackage[vcentermath]{youngtab}
\usepackage{relsize}
\usepackage{stackengine}
\usepackage{datetime}

\usepackage{hyperref}
\usepackage{physics}
\usepackage[dvipsnames]{xcolor}
\usepackage{tikz}

\numberwithin{equation}{section}

\usepackage{verbatim}

\newcommand{\be}{\begin{equation}}
\newcommand{\ee}{\end{equation}}

\newcommand{\bml}{\begin{multline}}
\newcommand{\emll}{\end{multline}}
\newcommand{\eqfig}[2]{\vcenter{\hbox{\includegraphics[width=#1]{#2}}}}

\newcommand{\nn}{\nonumber}

\def\({\left(} \def\){\right)}
\def\[{\left[} \def\]{\right]}

\def\Im{\text{Im}}

\def\da{a^{\dagger}}

\def\mO{\mathcal{O}}
\def\mC{\mathcal{C}}

\def\eps{\epsilon}

\def\v{\vec}

\def\g{\gamma}

\def\lam{\lambda}

\def\d{\partial}

\def\o{\omega}

\newcommand{\la}{\langle}
\newcommand{\ra}{\rangle}

\newcommand{\bea}{\begin{eqnarray}}
\newcommand{\eea}{\end{eqnarray}}

\usepackage[left=2cm,right=2cm,top=2cm,bottom=2cm]{geometry}
\titleformat{\section}{\large\bfseries}{\thesection.}{4pt}{}
\titlespacing{\section}{0pt}{22pt}{6pt}

\titleformat{\subsection}{\normalfont\bfseries}{\thesubsection.}{4pt}{}
\titlespacing{\subsection}{0pt}{18pt}{6pt}

\titleformat{\subsubsection}{\normalfont\itshape}{\thesubsubsection.}{4pt}{}
\titlespacing{\subsubsection}{0pt}{16pt}{6pt}

\def\ie{\begin{equation}\begin{aligned}}
\def\fe{\end{aligned}\end{equation}}

\def\tilde{\widetilde}

\def\hat{\widehat}

\def\bar{\overline}

\def\d{\partial}

\def\grad{\nabla}

\def\1{{\mathds 1}}

\def\Im{\mathop{\rm Im}}

\def\mL{\mathcal{L}}

\def\o{\omega}
\def\v{\vec }

\DeclareFontShape{OT1}{cmr}{mx}{n}%
    {<->cmr10}{}
\newcommand{\mytitlefont}{\fontseries{mx}\selectfont}
\DeclareMathAlphabet{\titlemath}{OT1}{cmr}{mx}{n}

\newcommand{\bi}{\begin{itemize}}
\newcommand{\ei}{\end{itemize}}

\newcommand{\sss}{\subsubsection}

\usepackage{bm}

\begin{document}

\begin{titlepage}

\begin{center}

~\\[1cm]

{\fontsize{20pt}{0pt} \mytitlefont 
Beyond the Boltzmann equation for weakly coupled quantum fields}\\[10pt]

~\\[0.2cm]

{\fontsize{14pt}{0pt}Xu-Yao Hu{\small $^{1}$} and Vladimir Rosenhaus{\small $^{2}$}}

~\\[0.1cm]

 \it{$^1$ Institute for Advanced Study, Tsinghua University}, \\ \it{Beijing, China}\\[10pt]

\it{$^2$ Initiative for the Theoretical Sciences}\\ \it{ The   Graduate Center, CUNY}\\ \it{
 365 Fifth Ave, New York, NY}\\[.5cm]

~\\[0.6cm]

\end{center}

\noindent 

We study the kinetic theory of a weakly interacting quantum field. Assuming a state that is close to homogeneous and stationary, we derive a closed kinetic equation  for the rate of change of the occupation numbers, perturbatively in the coupling. For a dilute gas, this reproduces the quantum Boltzmann equation, which only accounts for two-to-two scattering processes. Our expression goes beyond this, with terms accounting for multi-particle scattering processes, which are higher order in the density.

\vfill

\newdateformat{UKvardate}{%
 \monthname[\THEMONTH]  \THEDAY,  \THEYEAR}
\UKvardate
\today\\
\end{titlepage}

\tableofcontents

\section{Introduction}

The Boltzmann equation is the backbone of kinetic theory, compactly describing the rate of change of the  density $n(x,\vec k_1,t)$ of particles with momentum $\vec k_1$ at position $x$ \cite{Liboff, Kolb}, 
\be \label{BE}
\(\frac{\d }{\d t} + \frac{\vec k_1}{m} {\vdot} \grad\) n_1 =  -\int\! d^d k_2 d^d k_3 d^d k_4 |\mathcal{A}|^2\(n_1 n_2 - n_3 n_4\)\, \delta^{d+1}(k_{12;34})~,
\ee
where $n_i \equiv n(x,\vec k_i, t)$, $d$ is the spatial dimension, the delta function enforces  both momentum and energy conservation, with $k_{12;34} \equiv k_1{+}k_2{-}k_3{-}k_4$, and $\mathcal{A}$ is the scattering amplitude. 
The equation is conceptually transparent: the density of  particles with momentum $k_1$  decreases  when a particle with momentum $k_1$ collides with another of momentum $k_2$, producing outgoing particles with momenta $k_3$, $k_4$. The probability of this process is proportional to the square of the scattering amplitude  multiplied by the  number of incoming particles at position $x$, and the process is summed over possible momenta of the ingoing and outgoing particles. The time-reverse process increases $n(x,\v k_1, t)$. 

The validity of the Boltzmann equation hinges on the assumption that the dominant interactions involve two-to-two scattering events with minimal overlap. Higher-order corrections, which scale with higher powers of the density, become particularly significant in low dimensional systems due to memory effects arising from multiple correlated collisions \cite{PhysRev.139.A1763,PhysRevLett.25.1257, dorfman2015nonequilibrium, Dorfman, PhysRevLett.121.176805}. Higher order terms will involve multi-particle scattering amplitudes \cite{green1956boltzmann, cohen1962generalization,zwanzig1963method, bogoliubov1960problems,brocas1967comparison, balescu1960irreversible, Prigogine}, and it is challenging to write a systematic, simple, and useful higher order equation. 

At first pass, the quantum Boltzmann equation is  a simple modification of its classical counterpart. Its accounts for the quantum attraction/repulsion of bosons/fermions, respectively, by adding a factor of $1\pm n_i$ for the outgoing particles,
\be \label{QBE}
\!\!\!\!\!\Big(\frac{\d }{\d t} + \frac{\vec k_1}{m} {\vdot} \grad\Big) n_1 ={-}\!\!\int\! d^d k_2 d^d k_3 d^d k_4 |\mathcal{A}|^2\Big( n_1 n_2(1{\pm} n_3)(1{\pm}n_4)- (1{\pm}n_1) (1{\pm}n_2)n_3 n_4\Big)\, \delta^{d+1}(k_{12;34})
\ee
Of course, this equation  assumes a quasi-classical limit; quantum particles do not have a well-defined position and momentum. Concretely, it  assumes that the state is close to homogenous in both space and time (close to stationary), in order to minimize the impact of the uncertainty relation.

 More formally, a standard derivation of the quantum kinetic equation begins with the Schwinger-Dyson equations for the Green's function and self-energy on a Keldysh contour. These equations are inherently bilocal in space and time. To transform them into a closed equation for the Green's function $G(x_1, x_2)$, one typically works in the weak-coupling regime. Achieving a local form like (\ref{QBE}) requires a quasi-classical approximation, which formally involves a gradient expansion in terms of $x_1 {+} x_2$ and the momentum, defined as the Fourier transform of $x_1{ - } x_2$.

In this paper we assume, at the outset, a Hamiltonian with weakly coupled interactions and  a state that is spatially homogenous, nearly stationary, and close to Gaussian. The latter assumption (on the state) allows for a  quantum kinetic equation that is  manifestly local in time. The former assumption (on the coupling) allows for  a   straightforward and systematic derivation of the higher order terms (in the coupling) in the quantum kinetic equation. The derivation will be no harder (and the result more conceptually transparent) than the derivation of the classical wave kinetic equation, found perturbatively in the nonlinearity for weakly interacting  waves in Ref.~\cite{RS1, RS2, RSSS}, which served as motivation for this work.
We will give simple rules for writing down contributions to the kinetic equation at any order in the coupling, and each term will have as intuitive an interpretation as the tree-level quantum Boltzmann equation (\ref{QBE}), but will now involve multiple scatterings. 

One way to derive the classical  wave kinetic equation is by utilizing the equations of motion to relate the occupation number, $n_k = \la a_k^{\dagger} a_k\ra$ to the equal-time four-point correlation function \cite{Falkovich}:
\be \label{11eq}
\frac{\d n_1}{d t} = 4\Im  \int d^d k_2 d^d k_3 d^d k_4\, \delta(\v k_{12;34})\lambda_{1234} \la a^{\dagger}_1 a_2^{\dagger} a_3 a_4\ra~,
\ee
where  $\lam_{1234}$ is a general, and potentially momentum-dependent, quartic interaction. The correlation function on the right-hand side is computed self-consistently in a state that is close to Gaussian, having occupation numbers $n_k$, and close to stationary. 

There are multiple choices for what is being averaged over in the classical correlation function on the right-hand side of (\ref{11eq}). The first is over the  phases of the $a_k$ for the modes not entering the correlator \cite{RS1, onorato2020straightforward}, analogous to the phase-space averaging common in deriving the Boltzmann equation. 
The second, which is the simplest at the technical level,  is to modify the equations of motion by adding a  Gaussian-random forcing and dissipation for each mode,  whose magnitudes are then taken to zero while maintaining a ratio set by the desired $n_k$ \cite{ZakharovLvov, RS2}. Physically, this artificially mimics the coupling that a mode has to the many other modes which act as a bath. A third averaging is to take the initial state of the field $a_k$ to be Gaussian, with variance $n_k$ \cite{deng2023, deng2024}. All three of these averaging procedures give the same results for the (late-time) kinetic equation. 

Our derivation of the quantum kinetic equation will start by reinterpreting (\ref{11eq}) as a quantum equation. The correlation function on the right-hand side will be computed perturbatively in the coupling using a path integral approach. Since it is an expectation value (also known as an in-in correlator), the  time in the path integral will run over a Keldysh contour. Averaging over Gaussian initial conditions will correspond to imposing boundary conditions on the Green's function. Alternatively, we can instead average over  infinitesimal random forcing combined with dissipation, which is accounted for by coupling the fields on the two branches of the Keldysh contour. To give a flavor of the higher-order terms: accounting for one-loop diagrams gives,
\bea \nn
\frac{\d n_1}{\d t} = 16\pi   \int d^d k_2 d^d k_3 d^d k_4\,  \lambda_{1234}^2 \Big((n_1{+}1)(n_2{+}1)n_3 n_4 -(n_3 {+}1)( n_4{+}1) n_1 n_2\Big)\\
\(1 + 2\mL_+ + 8\mL_-\) \delta(\o_{k_1k_2; k_3 k_4})\delta(\v k_{12;34})~, \label{QBE2}
\eea
where $\o_k$ is the frequency and $\mL_{\pm}$ are principal value integrals,
\small
\be  
\mL_+ =  2 \!\int\! d^dk_5 d^dk_6\frac{\lam_{5612}\lam_{3456}}{\lam_{1234}} \frac{n_5{+}n_6{+}1}{\o_{k_1 k_2;k_5 k_6}}\delta(\v k_{12;56})~, \ \ \ \ \ \mL_- = 2 \! \int\! d^d k_5 d^d k_6 \frac{\lam_{3516}\lam_{4625}}{\lam_{1234}} \frac{n_6{-}n_5}{\o_{k_1 k_6;k_3 k_5}}\delta(\v k_{16;35})~.
\ee
\normalsize
Throughout the paper we will assume the state is  homogenous --- if one wishes to generalize, the left-hand side can be replaced with  $\d_t \rightarrow \d_t + \vec{v}{\cdot} \v \nabla$, as is standard in the Boltzmann equation.

At weak coupling, the $\mL_{\pm}$ terms may be neglected, as they are of order $\lam$, and the scattering amplitude at leading order is $\lam_{1234}$, so this reduces to the standard quantum Boltzmann equation (\ref{QBE}), where  the spatial derivatives on the left-hand side are absent due to the assumption of homogeneity. If one interprets (\ref{QBE2}) as a quantum Boltzmann equation computed to higher order in the density, then the terms in $\mL_{\pm}$ involving $n_i$  represent such higher order terms, whereas the $1$ in $\mL_+$ contributes to a higher order in coupling correction to the scattering amplitude. Taking the large $n_i\gg 1$ limit, this equation reduces to the next-to-leading order classical wave kinetic equation \cite{RS1, RS2}.

The paper is organized as follows: In Sec.~\ref{sec2} we set up the Keldysh contour for computing correlation functions and derive the propagators. In Sec.~\ref{sec3} we compute equal-time correlation functions perturbatively in the coupling, which via the quantum version of (\ref{11eq}),  gives  the kinetic equation. This is done at tree level in Sec.~\ref{sec31} and at one loop in Sec.~\ref{sec32}. In Sec.~\ref{sec33} we give rules for writing the answer to arbitrary order. We conclude in Sec.~\ref{sec4}.  
Appendix~\ref{appendix:green-and-occupationnum} reviews properties of propagators on the Keldysh contour. Appendix~\ref{apC} presents a direct perturbative calculation of equal-time correlation functions.  Appendix~\ref{apB} extends the results in the main body of the paper to interactions that do not conserve particle number.

\section{Keldysh contour propagators} \label{sec2}
Consider the Hamiltonian of the standard real scalar field $\phi$ with a quartic interaction, in $d$ spatial dimensions,
\be\label{21}
H = \int d^d x  \Big( \frac{1}{2}(\d \phi)^2 +\frac{1}{2}m^2 \phi^2+ \frac{\lam}{4!} \phi^4\Big)~.
\ee
It is convenient to switch to momentum space, writing the field in terms of creation and annihilation operators,~\footnote{In order to simplify notation, in this expression, and in all others, the factor of $(2\pi)^d$ is not explicitly written. In other words, in all momentum integrals one should replace $d^d k \rightarrow \frac{d^d k}{(2\pi)^d}$, and similarly for the momentum conserving delta function, $\delta(\v k)\rightarrow (2\pi)^d \delta(\v k)$. }
\be \label{phi}
\phi(x) = \int \frac{d^d k}{\sqrt{2 \omega_k}} \( a_k e^{i k{\cdot} x} + a_k^{\dagger} e^{- i k{\cdot} x}\)~.
\ee
Since we will be interested in the kinetic equation -- which describes the rate of change of the occupation number -- it is more natural to work  with $a_k$ instead of $\phi_k$. In particular, our propagators and interactions will be expressed in terms of $a_k$ and $a_k^{\dagger}$.  
Inserting  $\phi$ into the Hamiltonian, the interaction has terms with varying numbers of creation and annihilation operators, 
\small
\be \label{23}
\int \!\!d^d x \, \phi(x)^4 =\!\! \int \!\prod_{i=1}^4 \frac{d^d k_i}{ \sqrt{2\o_{k_i}}} \! \Big(6 \da_{k_1} \da_{k_2} a_{k_3} a_{k_4} \delta(\v k_{12;34}) + (4\da_{k_1} a_{k_2} a_{k_3} a_{k_4} \delta(\v k_{1;234})+a_{k_1} a_{k_2} a_{k_3} a_{k_4}\delta(\v k_{1234})+ \text{h.c.})\Big)
\ee
\normalsize
where $\v k_{12;34}\equiv \v k_1 + \v k_2 - \v k_3 - \v k_4$, $\v k_{1;234} \equiv \v k_1{-} \v k_2{-} \v k_3{-}\v k_4$ and $\v k_{1234} \equiv \v k_1{+} \v k_2{+}\v k_3{+}\v k_4$. 
The terms with different numbers of creation and annihilation operators appear separately in the tree-level kinetic equation. For simplicity, we will focus on the first term, which conserves particle number. The extension to the particle number nonconserving terms is straightforward, and is  discussed in Appendix.~\ref{apB} at tree-level, and  one-loop level for $\lam \phi^4$ in Appendix~\ref{sec:phi4}. However, the rules for evaluating a general loop diagram that we formulate Sec.~\ref{sec33} are natural specifically for a Hamitlonian with a definite number of creation and annihilation operators; for earlier studies for the full $\phi^3$ and $\phi^4$ interactions see \cite{old1,old2,old3,old4}. 

Sticking to interactions that conserve particle number, we can just as well consider an arbitrary quartic interaction,  $\lam_{k_1 k_2 k_3 k_4}\equiv \lam_{1234}$, that is some function of the momenta $k_i$, 
 \be
 H = \int\! d^d k\, \o_k\, \da_k a_k + \int \prod_{i=1}^4 d^d k_i\, \lam_{k_1k_2k_3k_4} \da_{k_1} \da_{k_2} a_{k_3} a_{k_4}\delta(\v k_{12;34})~,
 \label{Hamiltonian}
 \ee
 where $a_k(t)$ is the Fourier mode of the field and $\o_k$ is the dispersion relation. We will assume that $\lam_{1234}$ is real, $\lam_{1234}=\lam_{3412}$; the results easily generalize to complex interactions. For the remainder  of the main body of the paper, we  work with the Hamiltonian (\ref{Hamiltonian}). The equations of motion are, as usual,  $ \dot a_k =-i \frac{\d H}{\d \da_k}$, and the Lagrangian is,
 \be
 L = -i\int\! d^d k\, \dot{a}_k^{\dagger} a_k - H~.
 \ee

\begin{figure}
\centering
 \includegraphics[width=0.45\columnwidth]{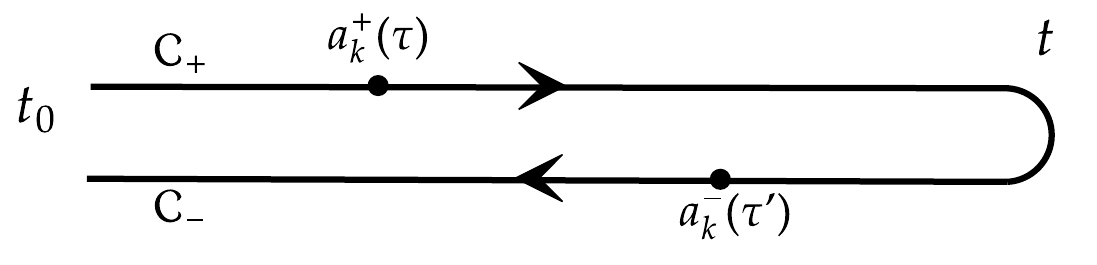}
\caption{For a path integral to compute expectation values of operators at time $t$, time should run along the Keldysh contour, starting and ending at $t_0$. } \label{FigKeldysh}
\end{figure}

We would like to compute expectation values of some operator $\mO(t)$ in a given state. The path integral formalism naturally computes  ``in-out'' correlation functions, in which operators are sandwiched between the state at time minus infinity and  time plus infinity, rather than the in-in correlation functions that we need. The path integral is easily adopted to  computing expectation values in a state  by using a Keldysh (folded contour). Specifically, 
 \be \label{23C}
 \la \mO(t)\ra =  \int_{\mathcal{C}} D a_p D \da_p\, \mO(t) e^{i S}~,
 \ee
where the contour $\mC$ runs from some initial time $t_0$ up to $t$, and then back to $t_0$, see Fig.~\ref{FigKeldysh}.
 
In Keldysh theory, see e.g. \cite{kamenev, Calzetta, Arnold:1998cy},  it is common to double the number of fields, with one on the upper branch of the contour and one on the lower branch; this makes it easier to account for the contour being folded. We will call these $a_k$ fields $a_k^+$ and $a_k^-$, where $a_k^+$ is on the upper branch (with time running the usual way) and $a_k^-$ is on the lower branch, with time running backwards. 
 We write 
 \be
 a_k(t) = a_k^+(t) + a_k^-(t)~,
 \ee
 where $a_k^+$ is  nonzero only on the upper branch and $a_k^-$ is  nonzero only on the lower branch. Since contour time runs backward along the lower branch, we pick up a minus sign from flipping time in the integration measure, leading to the action:
  \be
 S = \int dt \Big[\int d^dk\,  \big( i  {a_k^+}^{\dagger} \dot{a}^+_k -i  {a^-_k}^{\dagger} \dot{a}^-_k\big) - H(a^+) + H(a^-)\Big]~.
 \ee 
As we will see in the next section, the propagator will mix the upper and lower branch fields due to the presence of time-derivative terms in the action.

\subsection*{Averaging over random forcing}

 Next, we introduce  forcing and dissipation. This is an intermediate, technical step  (both forcing and dissipation are set to zero at the end) that is a convenient way of placing the system in a particular state. Adding the forcing is straightforward, by linearly coupling the field to some forcing function $f_k$. Dissipation, on the other hand, cannot usually be accounted for in a classical action without introducing auxiliary degrees of freedom. In the Keldysh formalism, the required ``auxiliary'' degrees of freedom naturally appear, and we simply need to couple the fields on the upper and lower branch, see e.g. \cite{Galley:2012hx,Galley:2014wla}, 
 \be
 S \rightarrow S + \int d t \int\! d^dk \(i \g_k a_k^{-}  {a_k^+}^{\dagger}  +if_k^*( a_k^+- a_k^- )\) + \text{c.c.}~.
 \ee
With this choice of action, the equations of motion are:
 \bea
  \dot a_k^+ &=&-i \frac{\d H(a^+)}{\d {a^+_k}^{\dagger} } -  \g_k {a^-_k} + f_k\\
  \dot a_k^- &=& -i\frac{\d H(a^-)}{\d {a^-_k}^{\dagger} } -  \g_k {a^+_k} +  f_k~.
 \eea
 In the classical limit, the fields on the upper and lower branch become equal,  $a_k^- = a_k^+$, and these are just the classical equations of motion in the presence of linearly coupled forcing and dissipation. 
 
 Like in the classical case, we will take the forcing to be Gaussian-random with variance $F_k$, 
 \be 
 P\[f\] \sim \exp\( - \int\! d t \int d^dk \frac{|f_k(t)|^2}{F_k}\)~, \ \ \ \  \la f_k(t) f_{p}^*(t')\ra = F_k \delta(\v k{-}\v p) \delta(t{ -} t')~.
 \ee
Integrating out the forcing leaves us with the action,
  \be \label{210}
 \!\!\!\!S = \!\int \!dt \Big[\int\! d^dk\, i\( ( {a_k^+}^{\dagger} \dot{a}^+_k - {a^-_k}^{\dagger} \dot{a}^-_k)+\g_k ( {a_k^+}^{\dagger} a_k^{-} - {a_k^-}^{\dagger} a_k^{+}   )+F_k  |a_k^+{-} a_k^-|^2\)  - H(a^+) + H(a^-) \Big]~.
 \ee

 It is useful to work with fields that are the sum and difference of the fields on the upper and lower branches,
\be
 A_k = \frac{1}{\sqrt{2}} (a_k^+ + a_k^-)~, \ \ \ \  \ \ \ 
    \eta_k = \frac{1}{\sqrt{2}} (a_k^+ - a_k^-)~.
  \label{Keldysh-rotated fields}
\ee
As we will see shortly,  $A_k$ can be regarded as the classical field and $\eta_k$ as the quantum field. Rewriting the Lagrangian (\ref{210}) and splitting it into a free and interacting part, $L = L_{\text{free}} +L_{\text{int}}$, gives:
\bea\nn
 \!\!\!\!\!\!\!\!  \!\!\!    L_{\text{free}} &=&i \int d^dk\, \( \eta_k^{\dagger}(\d_t {+} i \o_k {+} \g_k) A_k + A_k^{\dagger}(\d_t {+} i \o_k {-}\g_k) \eta_k
    +2 F_k |\eta_k|^2\) \\
 \!\!\!\!  \!\!\! L_{\text{int}} 
    &=&\!\! {-} \int \prod_{i=1}^4 d^d k_i\,  \delta(\v k_{12;34}) \lambda_{1234} \Big(\big(\eta_1^{\dagger} A_2^{\dagger} A_3 A_4 + A_1^{\dagger} A_2^{\dagger} \eta_3 A_4 \big)+\big(
    A_1^{\dagger} \eta_2^{\dagger} \eta_3 \eta_4
    + \eta_1^{\dagger} \eta_2^{\dagger} \eta_3 A_4
    \big)\Big)~. \ \ \ \
    \label{Lint}
\eea
The first interaction term, which contains one $\eta$ field, is classical. The second interaction term, which contains three $\eta$ fields, is quantum. 
Indeed, if we were working in a classical theory,  $\eta$ would appear linearly, as a Lagrange multiplier enforcing the classical equations of motion \cite{RS1}, within what is commonly called the Martin-Sigga-Rose formalism\cite{MSR, MSR2, MSR3}. 

\subsection*{Propagators, vertices,  Feynman rules}

It is simple to find the propagators, by inverting the quadratic (free) part of the action. As there are two fields ($A$ and $\eta$), there are four possible propagators. The $\eta$ two-point function  vanishes identically,~\footnote{This fact is independent of the dynamics, and is related to the BRST symmetries inherent to the Schwinger-Keldysh path integral, where $\eta$ is BRST exact \cite{Haehl:2016pec}.} leaving us with three propagators, which we represent diagrammatically as,
\begin{align}
    G^K_{k}(t_1,t_2) &= \la{A_k(t_1) A_k^{\dagger}(t_2)}\ra_{\mathcal{C}}
    =\eqfig{0.3 \columnwidth}{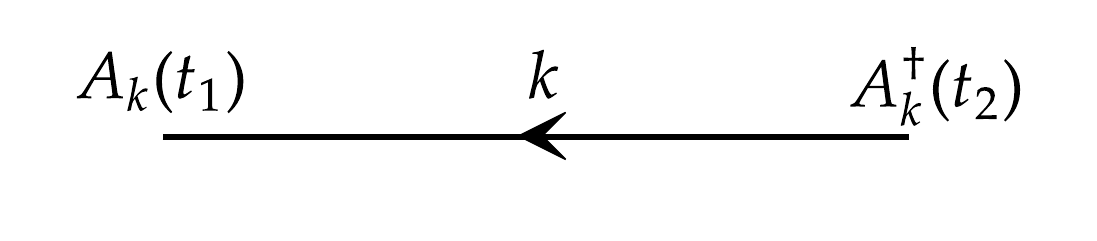} 
    \label{fig:GK-v2}
    \\
    G^R_{k}(t_1,t_2) &= \la{A_k(t_1) \eta_k^{\dagger}(t_2)} \ra_{\mathcal{C}}
    = \eqfig{0.3 \columnwidth}{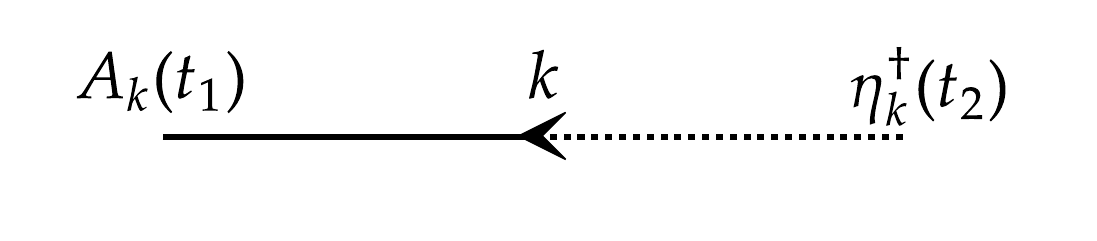} 
    \label{fig:GR-v2}
    \\
    G^A_{k}(t_1,t_2) 
    &= \la{\eta_k(t_1) A_k^{\dagger}(t_2)} \ra_{\mathcal{C}}
    =\eqfig{0.3 \columnwidth}{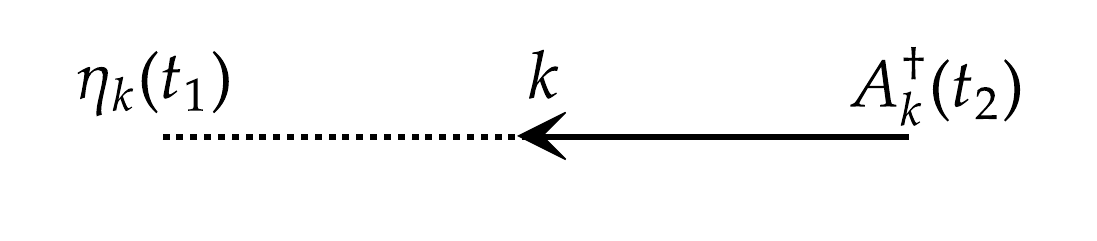} 
    \label{fig:GA-v2}
\end{align}
Here, the angle brackets indicate that the expectation values are evaluated in the initial state, 
and the subscript `$\mathcal C$' emphasizes that
the operators in the expectation value are contour-ordered, see Appendix \ref{appendix:green-and-occupationnum}.

Explicitly the propagators, in both frequency and time space, are
\bea
  \!\!\!\!\!\!  G^K_{k,\o}&=& \frac{2 F_k}{(\o-\o_k)^2 + \g_k^2}~,
    \ 
    \ \ \ \ \ \ 
    \ 
    G^K_k(t_1,t_2)= \frac{F_k}{\g_k} e^{- i \o_k t_{12} - \g_k |t_{12}|}~,
     \ \ \ \ \ \ \ 
    \label{GK}
    \\
    G^R_{k,\o}& =& \frac{i}{\o - \o_k + i \g_k}~,
    \ 
    \ \ \ \ \ \ \  \ \ 
    \
    G^R_k(t_1,t_2) = e^{-(i \o_k+\g_k)t_{12}} \theta(t_{12}) ~,
    \label{GR}
    \\
    G^A_{k,\o} &=& \frac{i}{\o - \o_k - i\g_k}~,
    \
    \ \ \ \ \ \  \ \  \
    \
    G^A_k(t_1,t_2) = -e^{-(i \o_k-\g_k)t_{12}} \theta(-t_{12})
    \ ~.
    \label{GA}
\eea
where we use the shorthand  $t_{12}\equiv t_1 -t_2$.
As is now clear, $G^K$ is the Keldysh Green's function, while $G^R$ and $G^A$ are the retarded and advanced Green's functions, respectively. 
The occupation number $n_k = \la \da_k(t) a_k(t)\ra$ of mode $k$  can be expressed in terms of the three Green's functions at equal time (see Appendix \ref{appendix:green-and-occupationnum}),
\be
    n_k = \lim_{t_2\to t_1 } \frac{1}{2}\left[
    G^K_k(t_1,t_2) + G^A_k(t_1,t_2) - G^R_k(t_1,t_2)
    \right] 
       =\frac{F_k}{2\gamma_k} - \frac{1}{2}~.
    \label{occupation-num-F}
\ee
We will be taking both $F_k$ and $\gamma_k$ to zero, $F_k, \g_k \rightarrow 0$, while maintaining constant $n_k$.
In this limit we may simplify the Keldysh Green's function (\ref{GK}), 
\be \label{GK2} 
G^K_{k, \o} = (2n_k + 1) \delta(\o{-}\o_k)~, \ \ \  \ \ \ \ \ \ \ G^K_k(t_1, t_2) = (2n_k+1) e^{- i \o_k t_{12}}~.
\ee

\begin{figure}[t]
    \begin{align} \nn
        \eqfig{0.35 \columnwidth}{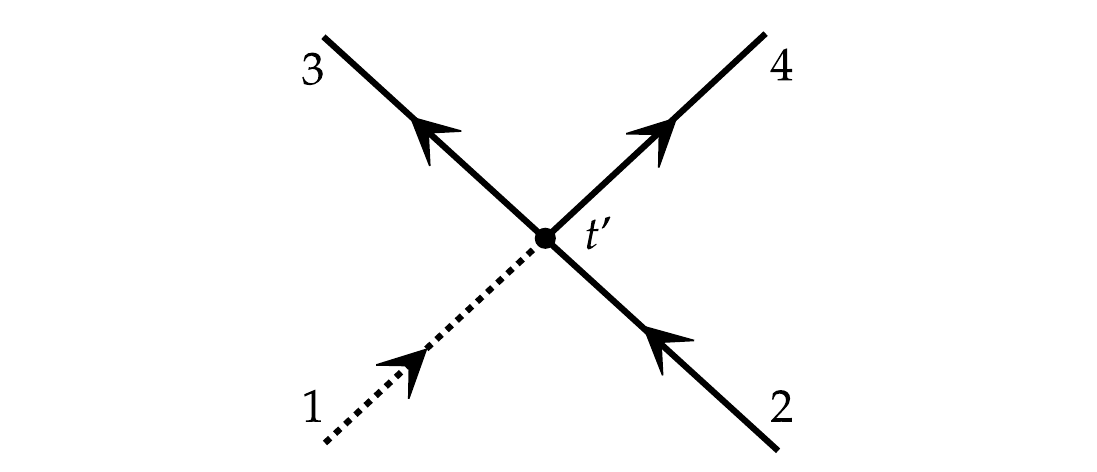} 
        \!\!\!\!\!\!\!\!\!\!\!\!\!\!\!\!\!\!\!\!\!\!\!\!\!\!\!\!\!
        \eqfig{0.35 \columnwidth}{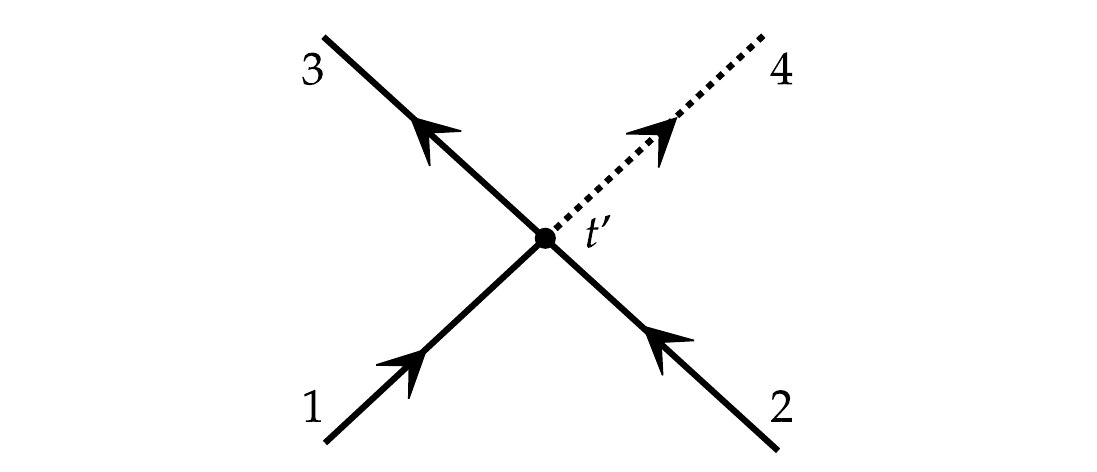} 
        \!\!\!\!\!\!\!\!\!\!\!\!\!\!\!\!\!\!\!\!\!\!\!\!\!\!\!\!\!
        \eqfig{0.35 \columnwidth}{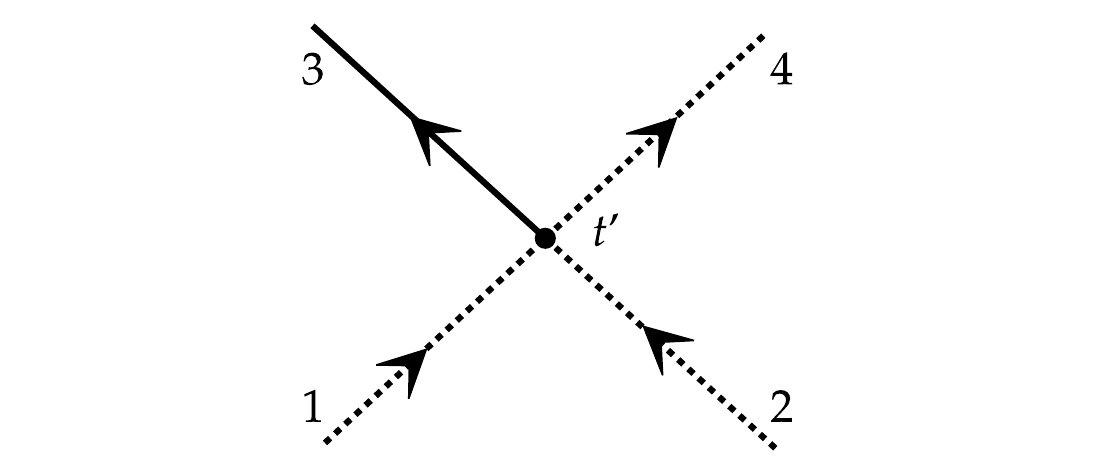} 
        \!\!\!\!\!\!\!\!\!\!\!\!\!\!\!\!\!\!\!\!\!\!\!\!\!\!\!\!\!
        \eqfig{0.35 \columnwidth}{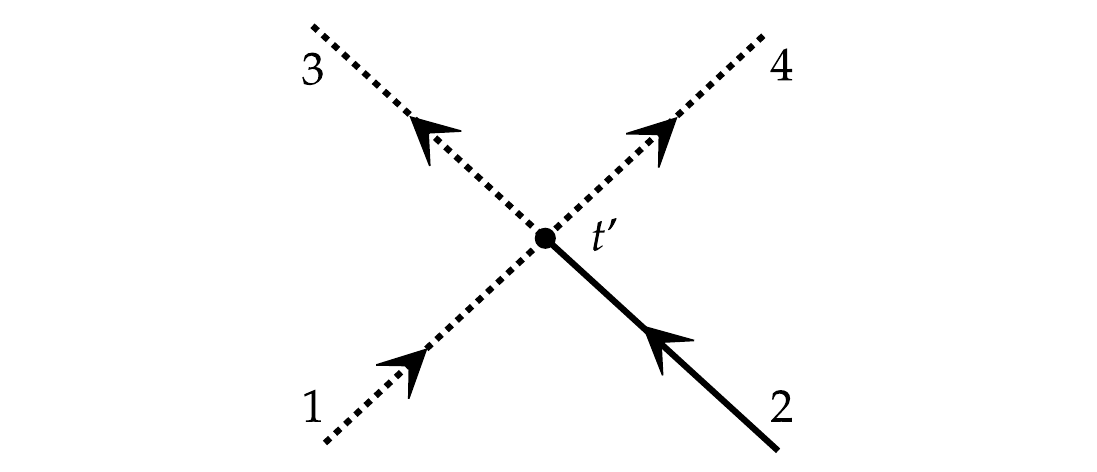}        
    \end{align}
    \caption{The first two vertices, which have one $\eta$ field and three $A$ fields, are classical. The last two vertices, which have one $A$ field and three $\eta$ fields, are quantum.}\label{vertices}
\end{figure}
Turning to the interacting part of the Lagrangian (\ref{Lint}), 
there are four types of vertices,  represented by the Feynman diagrams shown in Fig.~\ref{vertices}. The Feynman rule is to simply  assign a value of $- i \lam_{3412}$.

\subsection*{Averaging over initial conditions}
From looking at the form of the retarded (\ref{GR}) and Keldysh (\ref{GK2}) Green's functions, it is clear that we could have obtained the same result without ever introducing forcing and dissipation, by instead simply picking an appropriate Gaussian initial state. In particular,   in the absence of forcing and dissipation, the free part of the action takes the form $L_{\text{free}}$ in (\ref{Lint}) with $F_k =\g_k=0$,
\be
 S_{\text{free}} =i \int\! d^dk \int dt_1 dt_2 \begin{pmatrix}
        A_k^{\dagger}(t_1) & \eta_k^{\dagger}(t_1)
    \end{pmatrix}  \hat\Sigma_k(t_1,t_2)
    \begin{pmatrix}
        A_k(t_2) \\
        \eta_k(t_2)
    \end{pmatrix} 
   \ee
   where
   \begin{align}
        \hat  \Sigma_k(t_1,t_2)=\begin{pmatrix}
        0 & (\d_{t_1} + i \omega_k) \delta(t_1{-}t_2) \\
        (\d_{t_1} + i \omega_k) \delta(t_1{-}t_2) & 0
    \end{pmatrix}~.
   \end{align}
 The Green's function is given by the inverse, $ \hat{G}_k \hat{\Sigma}_k = \mathds{1}$. In terms of 
\begin{align}
    \hat G_k(t_1,t_2) = \begin{pmatrix}
        G_k^K(t_1,t_2) & G_k^R(t_1,t_2) \\
        G_k^A(t_1,t_2) & 0
    \end{pmatrix}
\end{align}
we see that $G^R$ and $G^K$  satisfy, 
\begin{align}
    -\d_{t_2} G^{R,A}_k(t_1,t_2) + i \omega_k G^{R,A}_k(t_1,t_2) &= \delta(t_1-t_2)~, \\
    -\d_{t_2} G^{K}_k(t_1,t_2) + i \omega_k G^{K}_k(t_1,t_2)  &= 0~,
\end{align}
which means that $G^R$ is given by (\ref{GR}), while $G^K$ takes the form, 
\be \label{GK2v2}
G^K_k(t_1, t_2) = g_k e^{- i \o_k t_{12}}
\ee 
with arbitrary $g_k$. In the path integral (\ref{23C}),  the precise Gaussian initial state at time $t_0$ can be incorporated through the choice of boundary conditions for $G_k(t_1, t_2)$ when  $t_1 {=} t_2{ =}t_0$, see e.g.~\cite{HR}. The choice  $g_k = 2n_k{+}1$ ensures that (\ref{GK2v2}) matches (\ref{GK2}).

While for a free theory one can pick any value of $n_k$, for an interacting theory the choice is strongly constrained. In particular, our derivation of the kinetic equation will assume that the state is close to stationary. One option is that the state is close to thermal. However, this is not the only option: for the (stationary) finite-flux solutions, upon obtaining the kinetic equation, as we will do in the next section, one finds the $n_k$ so that the collision term in the kinetic equation vanishes. In this sense,  $n_k$ is  determined a posteriori. 

\section{Correlation functions and the kinetic equation} \label{sec3}
The kinetic equation encodes the dynamics, describing how the occupation number of mode 
$k$ evolves over time. As shown in Appendix \ref{appendix:green-and-occupationnum}, this can be expressed in terms of 
the imaginary part of the equal-time four-point function, 
\be
 \!\!\  \!\!\!\!\!  \pdv{n_1}{t} = 4\Im \! \int \!\!\prod_{i=2}^4 d^d k_i \, \delta(\v k_{12;34})\lambda_{1234} \la a^{\dagger}_1 a_2^{\dagger} a_3 a_4\ra(t) 
    = \Im  \int \!\prod_{i=2}^4 d^d k_i \, \delta(\v k_{12;34}) \lambda_{1234} \big\la A_1^{\dagger} A_2^{\dagger} A_3 A_4\big\ra(t)
    \label{kineticEq-formal}
\ee
Here, the creation and annihilation operators in the middle expression belong to the original theory (\ref{Hamiltonian}), whereas the last correlation function is on the Keldysh contour and the $A$'s represent the Keldysh-rotated classical fields (\ref{Keldysh-rotated fields}).

\subsection{Tree-level kinetic equation} \label{sec31}

In this section, we compute the equal-time four-point function $ \la A^{\dagger}_1 A^{\dagger}_2 A_3 A_4 \ra(t)$   to leading order in the coupling, thereby reproducing the standard quantum kinetic equation. 
At tree level, each contributing diagram contains a  single vertex. One type of diagram involves three classical fields ($A$) and one quantum field ($\eta$) at the vertex, which we refer to as a classical vertex. For instance,
\begin{align}
    \eqfig{0.25 \columnwidth}{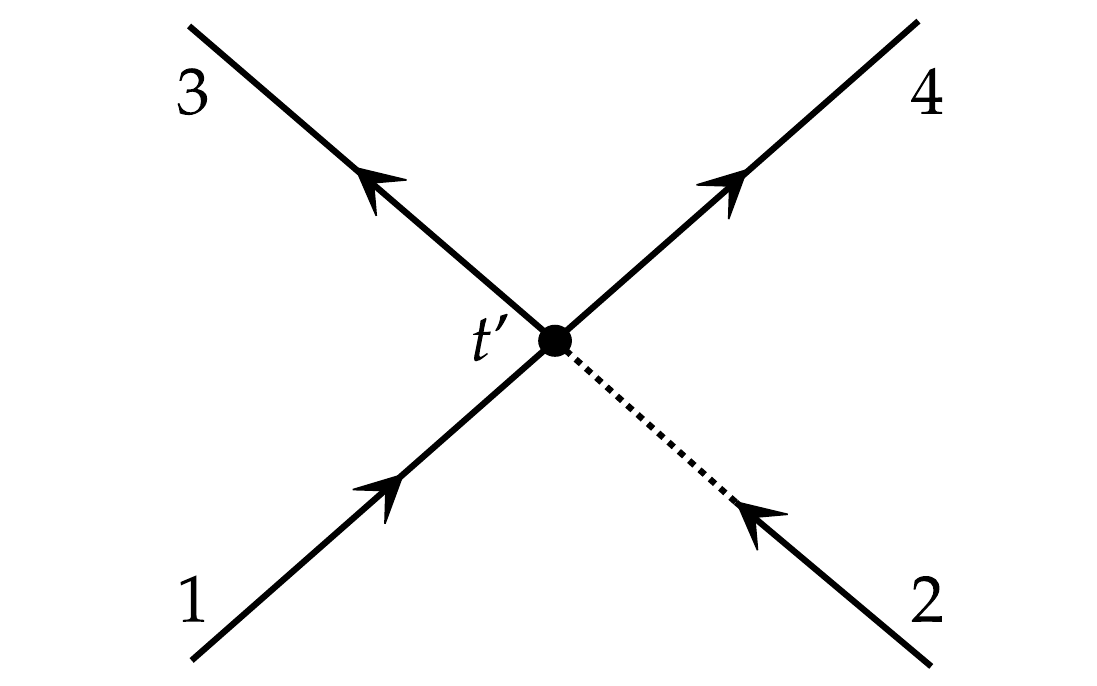}
    &= 
    -2i\lambda_{3412} \int dt' \ G^K_{k_1}(t', t) G^A_{k_2}(t',t)  G^K_{k_3}(t,t') G^K_{k_4}(t,t') 
    \notag\\
    &= 2\lambda_{3412} (2n_1{+}1) (2n_3{+}1)(2n_4{+}1) \frac{1}{\o_{34;12} {-} i\eps}~.
    \label{eq:tree-1}
\end{align}
The three other diagrams of this type immediately follow by choosing one of the other three legs attached to the vertex to be the $\eta$ field, denoted by the dashed line.

Another type of diagram involves three quantum and one classical field at the vertex, which we refer to as a quantum vertex. For instance,
\be
\!\!\! \!\!\! \!\!\! \!\!\!    \eqfig{0.2 \columnwidth}{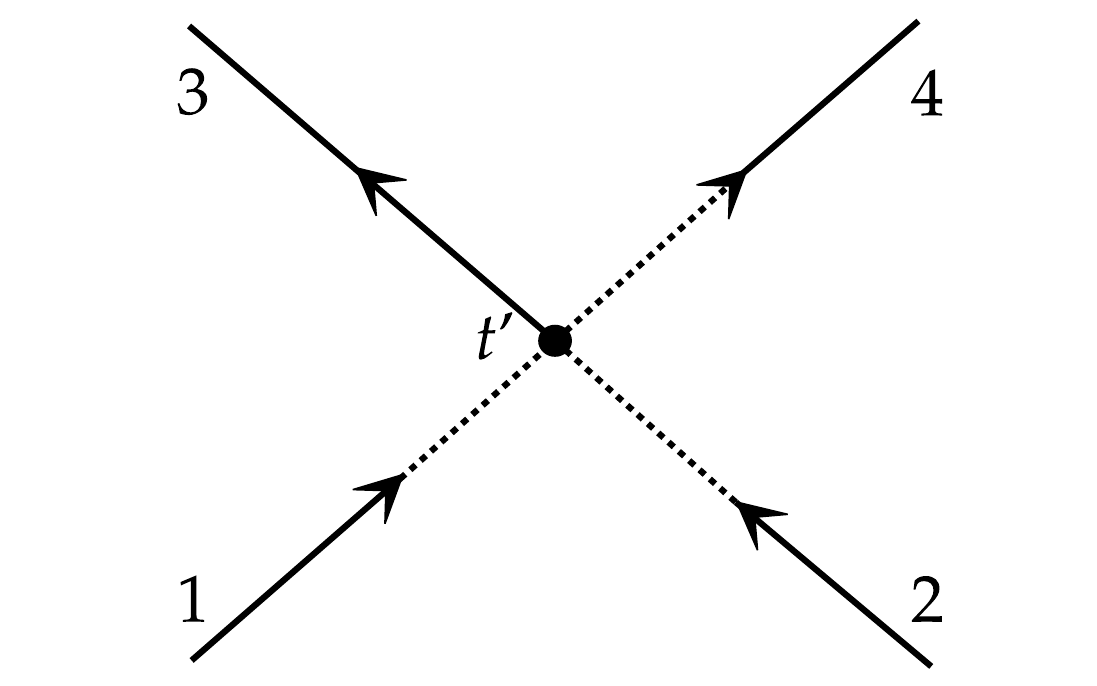}\!\!
    =
    -2i\lambda_{3412} \int dt'\ G^A_{k_1}(t',t) G^A_{k_2}(t',t) G^K_{k_3}(t, t') G^R_{k_4}(t,t') \!\!
    = -2\lambda_{3412} \frac{ (2n_3{+}1)}{\o_{34;12} {-}  i\eps}~.
    \label{eq:tree-2}
\ee
Similarly, three additional diagrams can be obtained by assigning one of the other legs to be the classical field, represented by the solid line. 
By summing all eight diagrams (four classical and four quantum) we obtain:
\be \label{34}
  \la A^{\dagger}_1 A^{\dagger}_2 A_3 A_4(t) \ra_{\text{tree}} = 16\lambda_{3412} \Big((n_1{+}1)(n_{2}+1)n_3 n_4-n_1 n_2(n_3{+}1)(n_4{+}1)  \Big)\frac{1}{\o_{34;12}{-}i\eps}~.
\ee
Explicitly, we used that, 
\be \label{35i}
N_1 N_2 N_3 N_4 \(\frac{1}{N_1}{+}\frac{1}{N_2}{-}\frac{1}{N_3}{-}\frac{1}{N_4} \) + (N_1 {+}N_2 {-}N_3{-}N_4) = 8  \Big((n_1{+}1)(n_{2}+1)n_3 n_4-n_1 n_2(n_3{+}1)(n_4{+}1)  \Big)~,
\ee
where $N_i \equiv 2n_i + 1$ and on the left-hand side the first term is from the classical vertices and the second term is from the quantum vertices. Using (\ref{kineticEq-formal}), we obtain the tree-level kinetic equation,
\be
    \pdv{n_1(t)}{t} = 16\pi \!\int \prod_{i=2}^4 d^d k_i\,  |\lambda_{1234}|^2 \big((n_1{+}1)(n_{2}+1)n_3 n_4-n_1 n_2(n_3{+}1)(n_4{+}1)  \big)\delta(\o_{12;34})\delta(\v k_{12;34})~.
    \label{tree-level-KE}
\ee
This is simply the standard quantum Boltzmann equation, where the scattering cross-section is the one at weak coupling. In the highly quantum regime $n_k\ll 1$ this becomes the classical Boltzmann equation for particles, whereas in the high occupation number $n_k \gg 1$ regime it becomes the classical kinetic equation for waves.

Returning to the integral appearing in (\ref{eq:tree-1}), we can  explicitly see the difference between the two types of averaging under consideration. The first, involving Gaussian random forcing and dissipation, means that the integral takes the form, 
\be
    \int_{t_0}^{t} d t' e^{-i( \o_{34;12} -i\eps)(t- t')} =\frac{ -i}{ \o_{34;12} {-}i\eps}~,
\ee
As a result of $\eps>0$, the contribution from the initial time $t_0$ doesn't enter. On the other hand, when averaging over a Gaussian initial state, there is no dissipation, so the expression instead takes the form,
\bea \nn
\int_{t_0}^{t} d t' e^{-i \o_{34;12}(t- t')} = \frac{ 1- e^{-i \o_{34;12} (t- t_0)}}{i \o_{34;12}}&=& -i\frac{1-\cos \o_{34;12}(t{-}t_0)}{\o_{34;12}}+ \frac{\sin \o_{34;12}(t{-}t_0)}{\o_{34;12}} \\
&\rightarrow& \frac{-i}{\o_{34;12}} + \pi \delta(\o_{34;12})~, \ \ \ \ \text{for} \ \ \ t{-}t_0\rightarrow \infty~.
\eea
In the second line, we took the late-time limit. In the imaginary component, the cosine oscillates rapidly and can therefore be dropped when integrated against a smooth function of time. The real part becomes a delta function, since, if $\o_{34;12}$ is not small, the numerator oscillates rapidly at late times and the expression vanishes, in the same sense. In short, whether averaging over Gaussian-random forcing and dissipation (with the variance of the former and the magnitude of the latter taken to zero) or over Gaussian initial conditions, the result is the same at late times.

\subsection{One-loop-level kinetic equation} \label{sec32}
We now look at the one-loop contributions to the equal-time four-point function, $\la A_1^{\dagger} A_2^{\dagger} A_3 A_4\ra$. Topologically, there is a single diagram, shown below. However, since we have two fields ($A$ and $\eta$), there are multiple variations of this diagram, depending on whether the propagators are Keldysh, retarded, or advanced. We begin with the $s$-channel diagrams.

A diagram with two classical vertices is,
\small
\begin{align} \nn
  &\ \ \  \ \ \ \ \ \ \ \ \ \ \ \ \ \  \ \ \ \ \ \ \ \ \ \ \ \ \ \ \ \ \ \ \  \ \ \ \ \ \ \ \   \eqfig{0.25 \columnwidth}{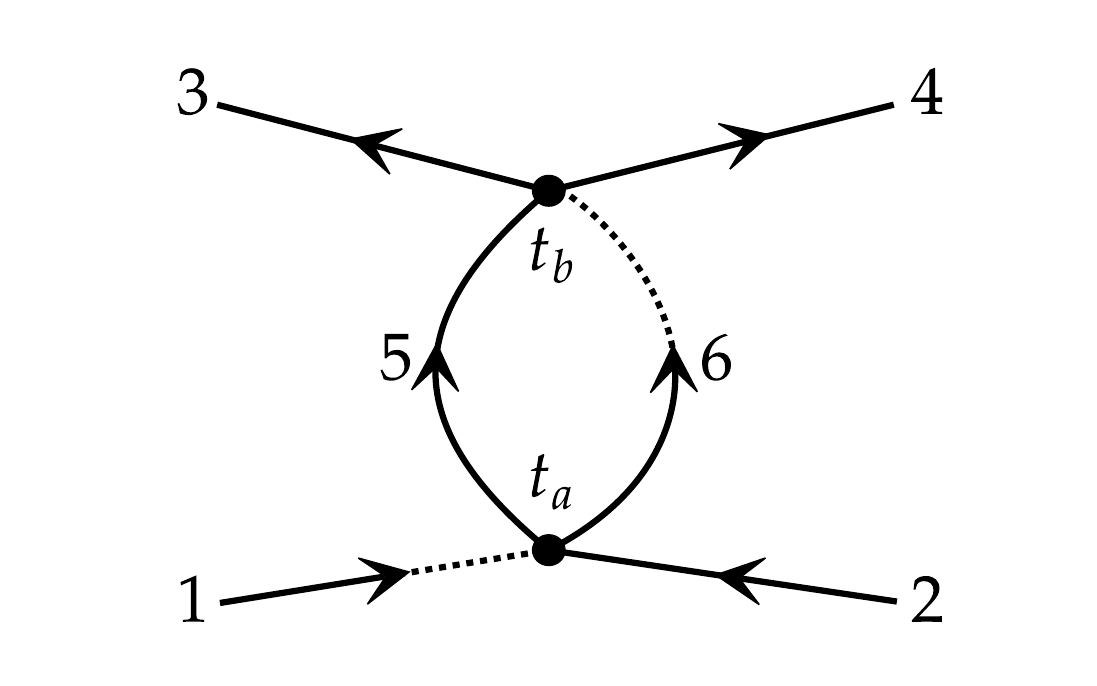}\\ \nn
&=  - \int\!\! d^d k_5 d^d k_6\, \delta(\v k_{12;56}) 
 \lam_{3456}\lam_{5612}   \int dt_a d t_b\ G^A_{k_1}(t_a,t) 
    G^K_{k_2}(t_a,t) G^K_{k_5}(t_b,t_a) G^A_{k_6}(t_b,t_a) G^K_{k_3}(t,t_b) G^K_{k_4}(t,t_b) 
    \\ 
    &= \frac{(2n_1 + 1)(2n_3+1) (2n_4+1)}{\o_{34;12}{-}i\epsilon}\int\!\! d^d k_5 d^d k_6\, \delta(\v k_{12;56}) 
    \lam_{3456}\lam_{5612} 
    \frac{(2n_5+1)}{\o_{34;56}{-}i\epsilon}
    ~.
    \label{s11-1}
\end{align}
\normalsize
where, in the second equality, we have  evaluated the time integrals, whose structure is identical to the classical case \cite{RSSS}. Another diagram with two classical vertices is, 
\bml\label{s11-2}
 \!\!\!\!\!\!\!\!   \eqfig{0.25 \columnwidth}{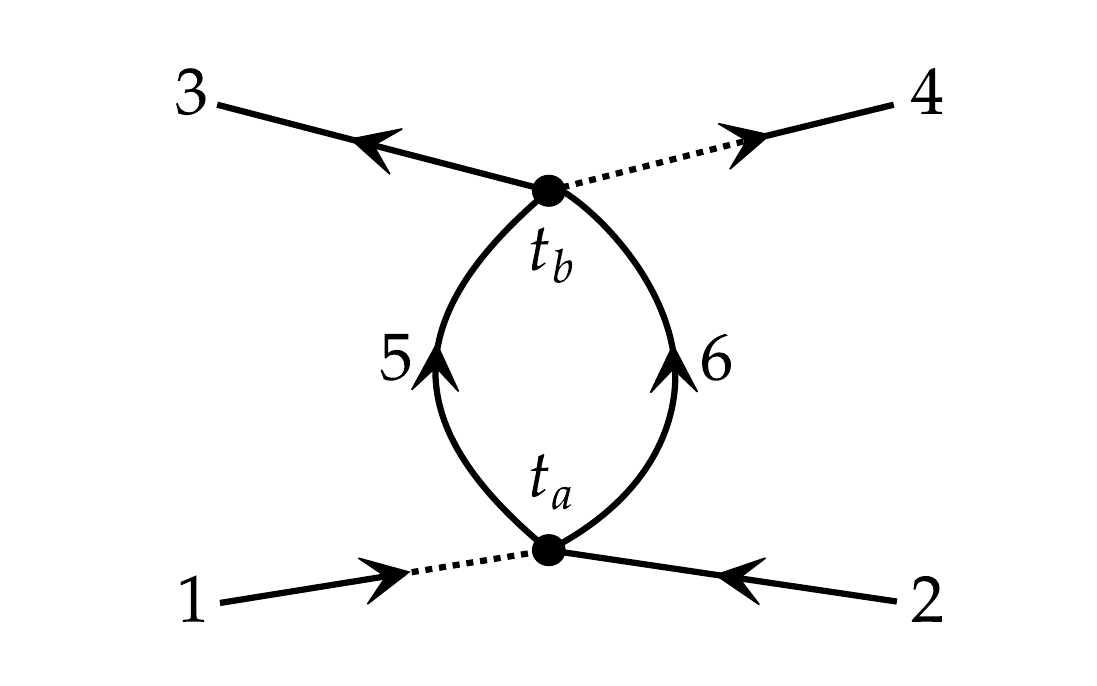}
    \!\!\!\!\!\!\!
    \!\!\! \!\!\!\!=
    - \frac{(2n_2{+}1)(2n_3{+}1)}{\o_{34;12} {-} i \eps}\!\!
    \int\!\! d^d k_5 d^d k_6\, \delta(\v k_{12;56})  \lam_{3456}\lam_{5612} (2n_5{+}1)(2n_6{+}1)\\[-25pt]
       \left[\frac{1}{\o_{56;12} { -} i\eps}
    +\frac{1}{\o_{34;56}{ -} i\eps}
    \right]~
     \ \ \ \ \  \
\end{multline}
There are also diagrams with one classical and one quantum vertex, such as,
\begin{align}
    \!\!\!\!\!\!\!
    \eqfig{0.25 \columnwidth}{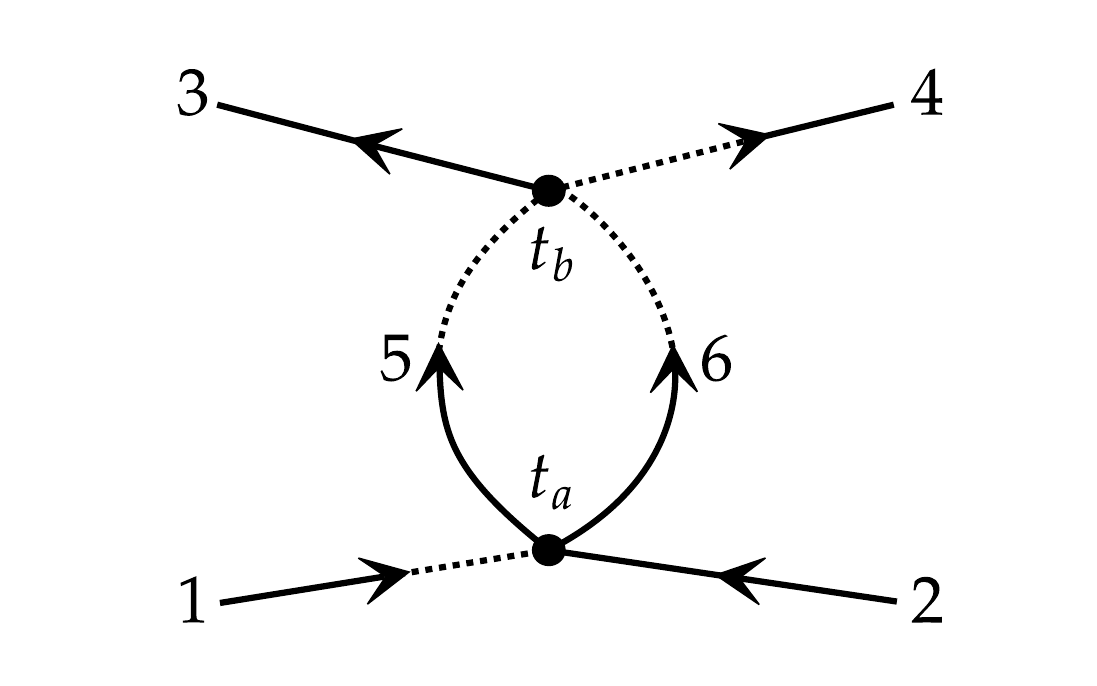}
    &= -\frac{ (2n_2{+}1) (2n_3 + 1)}{\o_{34;12} {-} i \eps}\int\!\! d^d k_5 d^d k_6\, \delta(\v k_{12;56})  \lam_{3456}\lam_{5612} 
   \frac{1}{\o_{34;56} {-} i \eps}~.
    \label{s13-1}
    \end{align}
    and 
    \begin{align}
    \!\!\!\!\!\!\!
    \eqfig{0.25 \columnwidth}{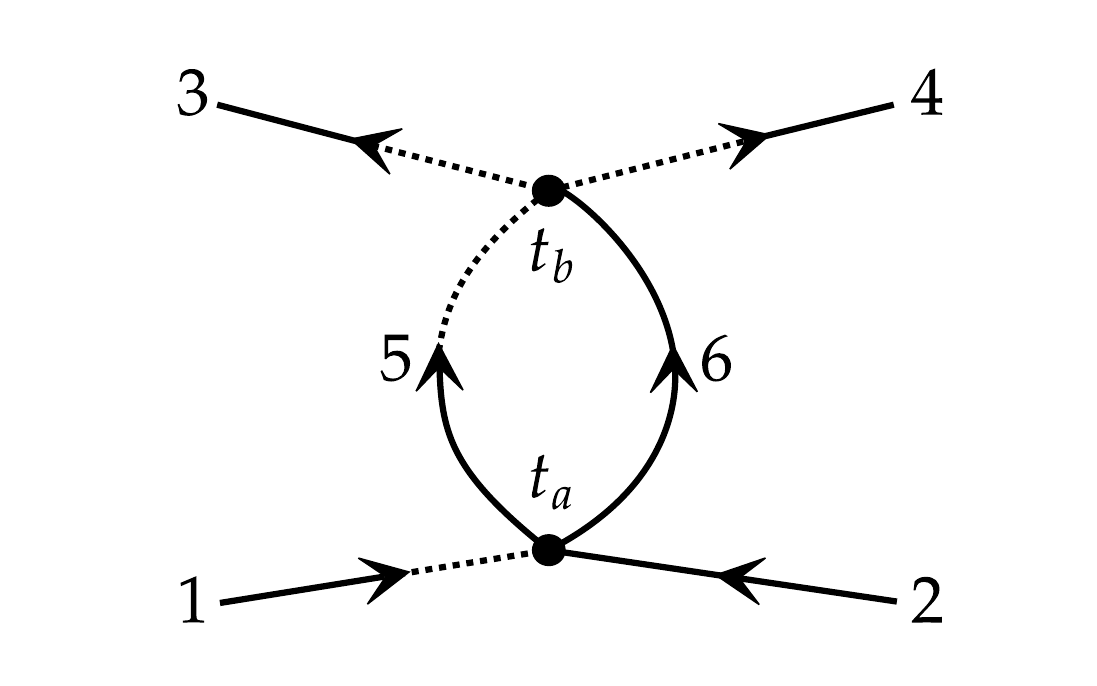}
    \!\!\!\!\!\!\!
    &=\frac{2n_4 {+} 1}{\o_{34;12} {-} i \eps}\int\!\! d^d k_5 d^d k_6\, \delta(\v k_{12;56})  \lam_{3456}\lam_{5612} \frac{2n_6{+}1 }{\o_{34;56} {-} i\eps}
    ~.
    \label{s13-2}
\end{align}

All  other diagrams can be derived from these four by exchanging the mode variables, and potentially complex conjugating. Note that 
a one-loop diagram involving two quantum vertices  vanishes identically: the presence of two quantum vertices forces the loop into one of the following two structures,
\be
    \!\!\!\!\!\!\!\!\!\!\!\!\!\!\!\!\!
    \eqfig{0.3 \columnwidth}{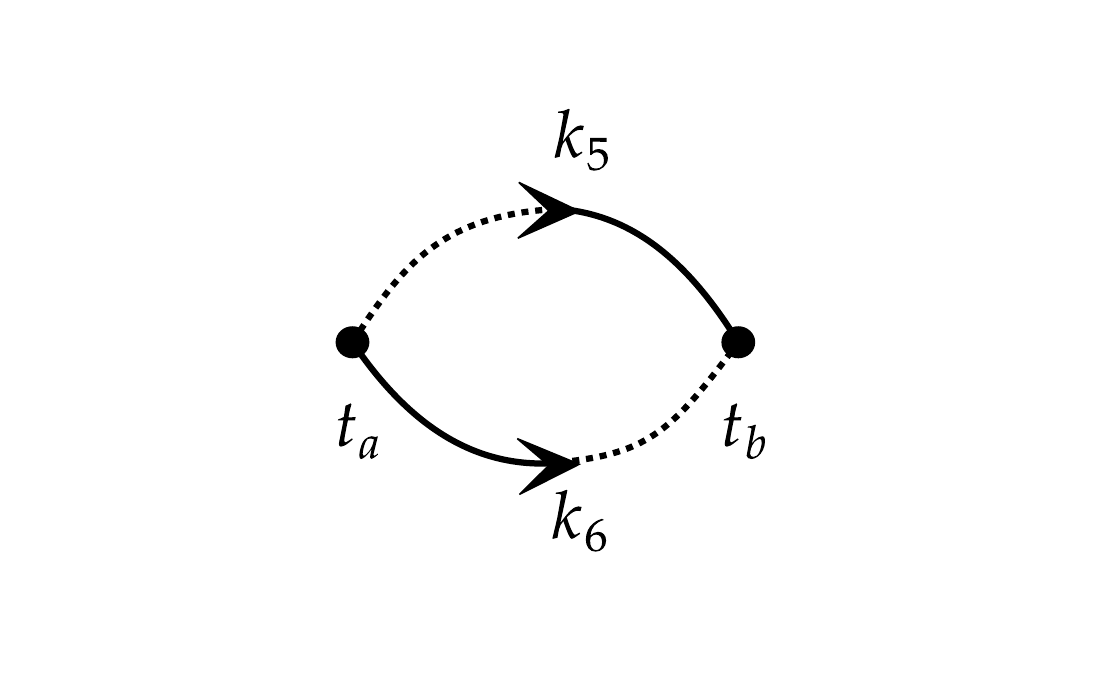}
    \!\!\!\!\!\!\!\!\!\!\!\!\!\!\!\!\!
    = G^R_{k_5}(t_b,t_a) G^A_{k_6}(t_b,t_a)  \ \ \ \ 
    \eqfig{0.3 \columnwidth}{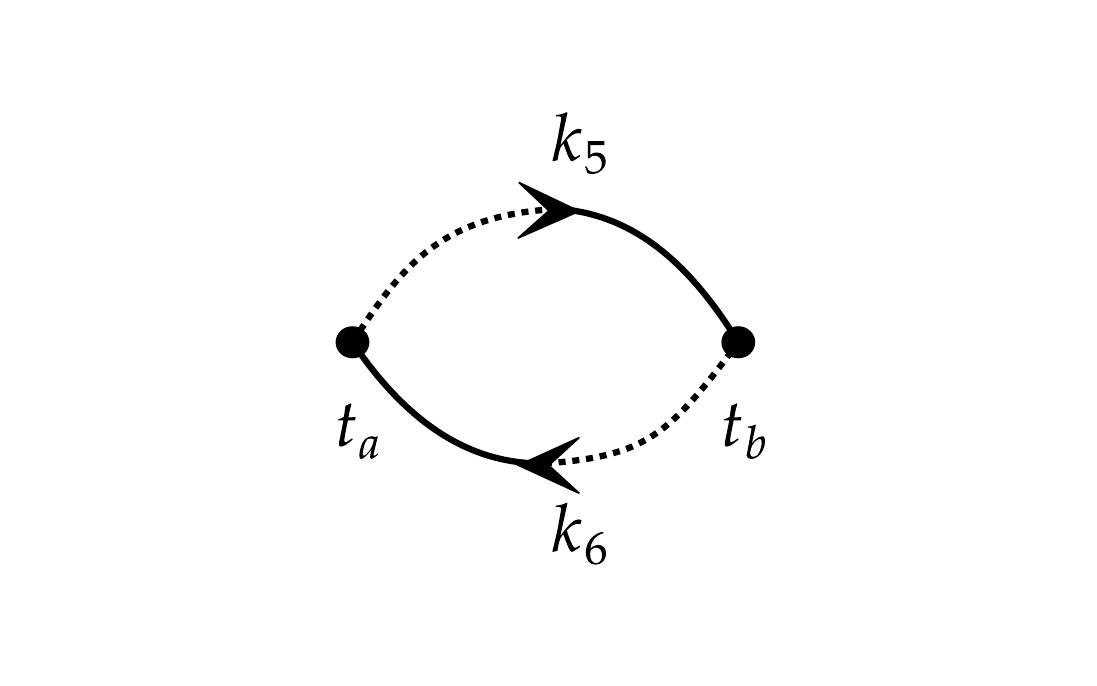}
    \!\!\!\!\!\!\!\!\!\!\!\!\!\!\!\!\!
    = G^R_{k_5}(t_b,t_a) G^R_{k_6}(t_a,t_b) 
    \ee
both of which have conflicting time-orderings of $t_a$ and $t_b$, $\theta(t_{ab})\theta(-t_{ab})$, forcing the integral to vanish.

Summing over the diagrams, the total $s$-channel contribution to the equal-time four-point function is 
\small
\begin{align} \nn
& \la A_1^{\dagger}A_2^{\dagger} A_3 A_4\ra_{s\text{-channel}}
    =\frac{16}{\o_{34;12} {-}i \eps}\int\!\! d^d k_5 d^d k_6\,   \lam_{3456}\lam_{5612}  \Big[ (n_3{+}n_4{+}1)\frac{n_1 n_2 (n_5{+}n_6{+}1) {-} n_5 n_6 (n_1{+}n_2{+}1)
    }{\o_{56;12} {-} i \eps} 
    \notag \\
    & \ \ \ \ \ \ \ \ \ \ \  \ \ \ \ \  \ \ \ \ \ \ \ \ \ \ \ \ + (n_1 {+} n_2{+}1) \frac{ n_3 n_4 ( n_5 {+} n_6{+}1) - n_5 n_6(n_3{+}n_4{+}1) }{\o_{34;56} {-}  i\eps}\Big]\delta(\v k_{12;56})
    \label{s-channel-total}
\end{align}
\normalsize
We also need to look at the one-loop diagrams in the $t$-channel. An example is
\be
    \eqfig{0.3 \columnwidth}{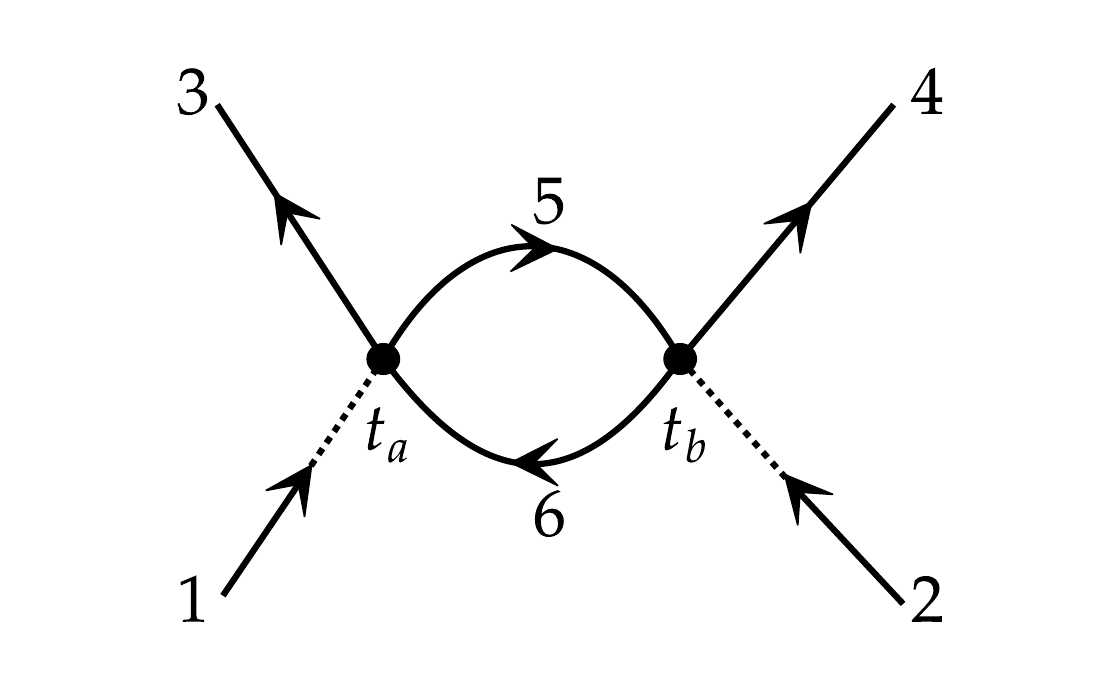}
\ee
where momentum conservation imposes $k_3 + k_5 = k_1 + k_6$. There are other $t$-channel diagrams with possible classical and quantum vertices, similar to those in the $s$-channel. It is simple to obtain the $t$-channel contribution from the $s$-channel result by a symmetry transformation \cite{RS2}, sending $6\to -6$, $3 \leftrightarrow -2$. This transforms the coupling  $\lam_{5612}\rightarrow \lam_{5-6 1 -3 } \equiv \lam_{3516}$, whereas  for the occupation numbers $i\rightarrow -i$ sends $n_i \rightarrow -1 - n_i$, and there is an overall sign of $-1$ for each arrow that is flipped ($3$ in this case). Accounting for the extra combinatorial factor of $2$, we get 
that the $t$-channel contribution is
\small
\begin{align}
 \la A_1^{\dagger}A_2^{\dagger} A_3 A_4\ra_{t\text{-channel}} 
    =&\frac{32}{\o_{34;12}{ -} i \eps} \int\!\! d^d k_5 d^d k_6\,\lambda_{3516} \lambda_{4625} \Big[ (n_2{-}n_4)\frac{n_1 n_6 (n_3{+}n_5{+}1) - n_3 n_5 (n_1{+}n_6{+}1)
    }{\o_{35;16}{ -} i \eps} 
    \notag \\
    +&(n_1 {-} n_3) \frac{ n_2 n_5 ( n_4 {+} n_6{+}1) {-} n_4 n_6(n_2{+}n_5{+}1) }{\o_{46;25} {-} i\eps}\Big]\delta(\v k_{16;35})~.
    \label{t-channel-total}
\end{align}
\normalsize
The $u$-channel  follows trivially by exchanging $3$ and $4$ in the $t$-channel contribution.

Thus, the quantum kinetic equation to order $\lam^3$ is, via (\ref{kineticEq-formal}), expressed in terms of the equal-time four-point function, which is the sum of the tree-level and one-loop diagrams, 
\bea \nn
    \pdv{n_1}{t} = 16\pi\,\int\prod_{i=2}^4 d^d k_i \lambda_{1234}^2 \Big((n_1{+}1)(n_{2}+1)n_3 n_4-n_1 n_2(n_3{+}1)(n_4{+}1) \Big)\\
    \Big(1 + 2\mL_+ + 8\mL_-\Big)\delta(\o_{12;34}) \delta(\v k_{12;34})~.
    \label{one-level-KE}
\eea
where
\be \label{Lpmin}
\mL_+ = 2\!\int\!\! d^d k_5 d^d k_6 \frac{\lam_{5612}\lam_{3456}}{\lam_{1234}} \frac{n_5{+}n_6{+}1}{\o_{12;56}  }\delta(\v k_{12;56})~, \ \ \ \ \mL_-= 2\!\int\!\! d^d k_5 d^d k_6  \frac{\lam_{3516}\lam_{4625}}{\lam_{1234}} \frac{n_6{-}n_5}{\o_{16;35} }\delta(\v k_{16;35})~.
\ee
Here, the $1/\omega$ denominators are understood as their principal values.
The $s$-channel  one-loop four-point function gives the $\mL_+$ term, while the $t$ and $u$ channels each give an $\mL_-$ term. In addition, here $\o_k$ is really the one-loop renormalized frequency, 
\be
\o_1 \rightarrow  \o_1 + 4\int\! d^d k_2\, \lam_{1212} (n_2+\frac{1}{2})~. 
\ee

\subsection{Higher loop diagrams} \label{sec33}
In this section, we provide a simple prescription for determining the contribution of any Feynman diagram to an equal-time correlation function. This serves as a quantum generalization of the rules presented in \cite{RSSS}. 

When evaluating a given Feynman diagram, it is useful to choose a definite time ordering of the internal vertices. Once this is done, the  time-dependent portion of the integrand is simply a product of exponentials $e^{-i \o_k t_{a b}}$ which connect vertices at times $t_a$ and $t_b$; the Keldysh, retarded, and advanced Green's function all have this same dependence, (\ref{GK}-- \ref{GA}). The result of the time integrals will be identical to that  discussed in \cite{RSSS}. Specifically, one moves from the earliest time to the latest time.~\footnote{We are using an opposite time ordering convention from \cite{RSSS}.} At each step, one draws an imaginary loop  enclosing all previously  visited vertices and writes down a factor
\be \label{step3}
\frac{1}{\o_{i} {+}\o_{j} {+}\ldots {-} \o_{a} {-} \o_{b}{-} \ldots - i\eps}~,
\ee
where $\o_i, \o_j, \ldots$ are the frequencies of all lines leaving this imaginary loop and $\o_a, \o_b, \ldots$ are the frequencies of all lines entering the imaginary loop. For example, for the $s$-channel diagram in the previous section, such as the one in (\ref{s11-1}),  the time ordering $t_a>t_b$ yields the factor,
\be \label{321}
\frac{1}{\o_{34;56}{ -}i\eps} \frac{1}{\o_{34;12} {-}i\eps} 
\ee
where the first term comes from the first imaginary loop, enclosing vertex  $b$, and the second term comes from the imaginary loop enclosing both vertices. This matches the $\o$ dependence written in (\ref{s11-1}), (\ref{s13-1}), (\ref{s13-2}), which, due to the advanced Green's function within the loop, must follow this time ordering.  One of the pieces in (\ref{s11-2})-- which, as a result of both Green's functions within the loop being Keldysh -- can have either ordering, $t_b>t_a$ or $t_b<t_a$. In fact,  rule (\ref{step3}) becomes evident when computing the expectation value directly in quantum mechanical perturbation theory, see Appendix~\ref{apC}. In particular, as one evolves the state from $t_0$ to $t$, each intermediate interaction vertex modifies the energy, by creating and annihilating particles. These $\o$ denominator factors reflect the total energy (in the free theory) at these intermediate times between $t_0$ and $t$. 

\begin{figure}\centering
\includegraphics[width=0.3\columnwidth]{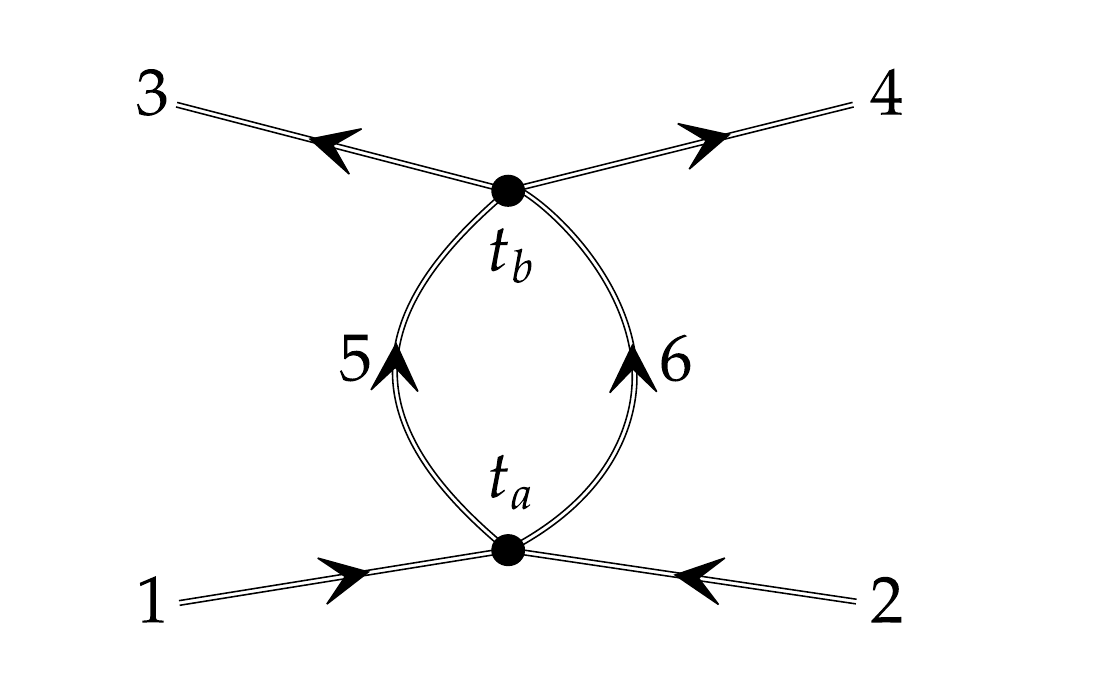}
\caption{The double line on the propagator indicates that it can be either a Keldysh Green's function (solid single line), or a retarded or advanced Green's function (solid/dashed line). This figure therefore includes all the $s$-channel diagrams, shown earlier in Figs.~\ref{s11-1} -- \ref{s13-2}. } \label{Figschan}
\end{figure}
In the previous section, for each vertex we had a choice of if it is a classical or a quantum vertex. In fact, we can consider all the  choices simultaneously. Let us redraw the $s$-channel diagram in (\ref{s11-1}),  being agnostic  if each vertex is classical or quantum, and correspondingly if the propagators are Kelydsh, retarded, or advanced, see Fig.~\ref{Figschan}.  Now consider the time ordering $t_a>t_b$. We look at vertex $a$, and see that we cannot have a dashed line coming out from $a$ in the direction of $b$. Since there are two lines going from $a$ to $b$, this means the vertex at $a$ must be classical, and that the one dashed line coming out of the classical vertex must be going in the direction of the external time. The other line going to the external time must therefore be the Keldysh propagator, so we get a factor of $n_1{+}n_2$. Moving now to vertex $b$,  any of the lines coming out of it can be dashed or solid, so the vertex can be either classical or quantum. In total we get the factor, 
\bml \label{322}
(N_1{+}N_2) \[  N_3 N_4 N_5 N_6\(\frac{1}{N_5}{+}\frac{1}{N_6}{-}\frac{1}{N_3}{-}\frac{1}{N_4} \) + (N_5 {+}N_6 {-}N_3{-}N_4) \] \\
= 16(n_1{+}n_2{+}1) \Big(n_3 n_4 (n_5{+}1)(n_6{+}1)-(n_3 {+}1)( n_4{+}1) n_5 n_6\Big)
\end{multline}
where $N_i \equiv 2n_i+1$, see (\ref{35i}). Combining (\ref{321}) with (\ref{322}), and the appropriate combinatorial factor and multiplying by the couplings, reproduces the second term in (\ref{s-channel-total}). 

The other option for the time ordering of the internal vertices is $t_a<t_b$. From the time integral we get the factor, 
\be  \label{323}
\frac{1}{\o_{56;12} {-}i\eps} \frac{1}{\o_{34;12} {-}i\eps} 
\ee
where the first term comes from the first imaginary loop, enclosing vertex at $a$, and the second term comes from the imaginary loop enclosing both vertices. The factor involving the occupation numbers is,
\bml\label{324}
-(N_3{+}N_4) \[  N_1 N_2 N_5 N_6\(\frac{1}{N_1}{+}\frac{1}{N_2}{-}\frac{1}{N_5}{-}\frac{1}{N_6} \) + (N_1 {+}N_2 {-}N_5{-}N_6) \] \\
= -16(n_3{+}n_4{+}1) \Big((n_1{+}n_2{+}1)n_5 n_6 -(n_5 {+} n_6{+}1) n_1 n_2\Big)
\end{multline}
where the first term, $N_3{+}N_4$, comes from vertex $b$ having to be classical. Combining (\ref{323}) and (\ref{324}) gives the first term in (\ref{s-channel-total}). \\[5pt]

We are now ready to give simple rules for writing down any equal-time correlation function to any order in the coupling. For a given Feynman diagram (drawn for the theory expressed in terms of the original variables -- only $a$ fields):
\\[5pt]
\noindent \textbf{Rules:}

\begin{enumerate}
\item Pick an ordering of the times at each vertex. The time $t$ at which the correlation function is being evaluated must be the largest time.
\item Start at the earliest time on the diagram and move from vertex to vertex in increasing order of their times, until finally reaching the vertex at the latest time. Each subsequent vertex must be adjacent to at least one previously visited vertex.  At each step in this process, draw an imaginary loop  enclosing all vertices visited so far. Write down a factor of 
\be
\frac{1}{\o_{i} {+}\o_{j} {+}\ldots {-} \o_{a} {-} \o_{b}{-} \ldots - i\eps}~,
\ee
where $\o_i, \o_j, \ldots$ are the frequencies of all lines leaving this imaginary loop and $\o_a, \o_b, \ldots$ are the frequencies of all lines entering the imaginary loop.

\item 
Start at the latest time on the diagram and move from vertex to vertex in decreasing order of their times, until finally reaching the vertex at the earliest time. Each next vertex must be a neighbor of at least one previously visited vertex.
At each step, when writing down the occupation number factor for each vertex only look at the lines coming out of the vertex that are going to another vertex that is at a later time. This makes the interaction at this vertex effectively $q$ body, if there are $q$ such lines.  Let $a$ be the index for the lines entering the effective vertex and $i$ the index for the lines leaving the effective vertex. Write down a factor of 
\be \label{325}
 \prod_a n_a \prod_{i}(1+ n_{i})- \prod_a(1+n_a)\prod_{i} n_{i}
\ee
This is the standard result that one expects for a tree-level interaction: it is weighted by the product of the occupation numbers $n$ of the ingoing particles, and by a factor of the product of $1+n$ for the outgoing particles (reflecting the Bose enhancement). If we had fermions instead of bosons we would simply replace this by $1-n$. The second term is the time reverse process. 

Explicitly, for quartic, cubic,  quadratic, and one-body effective interactions this factor is: 
\begin{align}
    \!\!\!\!\!\!\! \!\!\!\!\!\!\!
    \eqfig{0.3 \columnwidth}{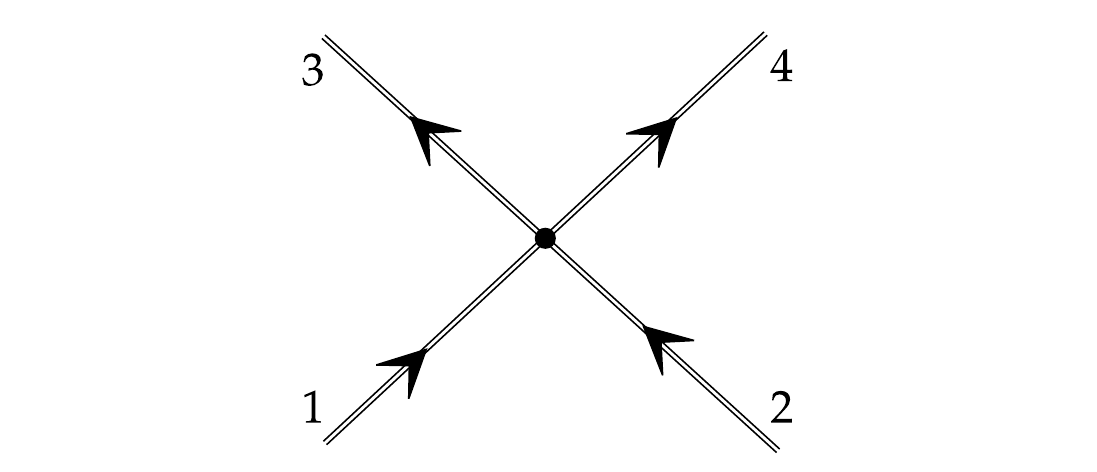}
    \!\!\!\!\!\!\! \!\!\!\!\!\!\!
    &= n_1 n_2(1{+}n_3)(1{+}n_4)-n_3n_4 (1{+}n_1)(1{+} n_2) 
    \label{4v}
    \\
    \!\!\!\!\!\!\! \!\!\!\!\!\!\!
    \eqfig{0.3 \columnwidth}{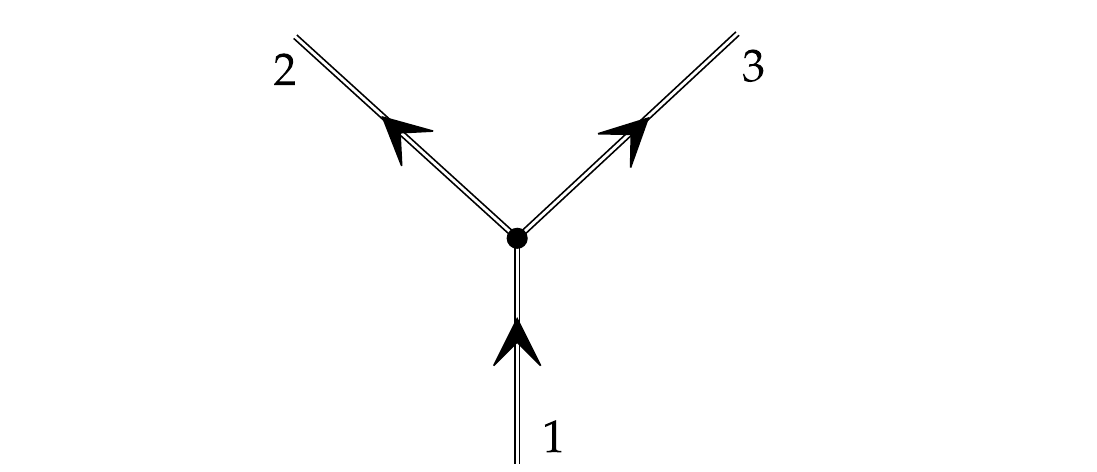}
    \!\!\!\!\!\!\! \!\!\!\!\!\!\!
    &=n_1 (1{+}n_2)(1{+}n_3) - (1{+}n_1)n_2 n_3
    \label{3v}
    \\[-7pt]
    \!\!\!\!\!\!\! \!\!\!\!\!\!\!
    \eqfig{0.3 \columnwidth}{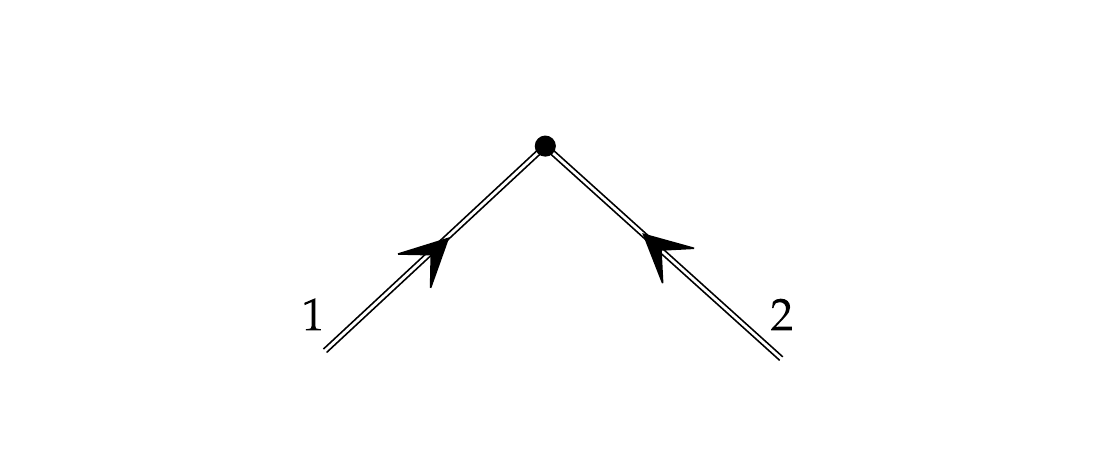}
    \!\!\!\!\!\!\! \!\!\!\!\!\!\!
    &=n_1n_2 - (1{+}n_1)(1{+}n_2) = - (1{+}n_1{+}n_2) 
     \label{2v1}
    \\[-15pt]
    \!\!\!\!\!\!\! \!\!\!\!\!\!\!
  \eqfig{0.3 \columnwidth}{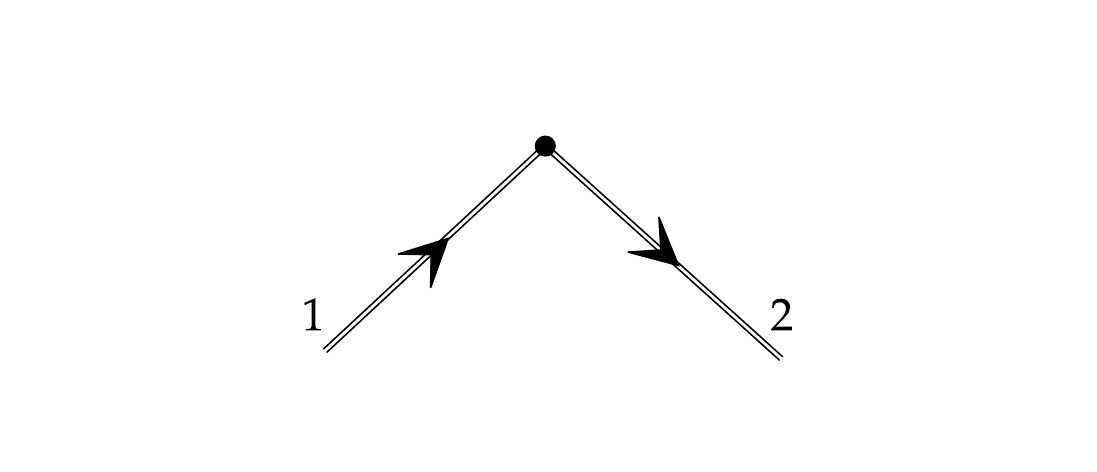}
    \!\!\!\!\!\!\! \!\!\!\!\!\!\!
    &=n_1(1{+}n_2) - (1{+}n_1) n_2 = n_1 {-} n_2  
      \label{2v2}
    \\[-15pt]
    \!\!\!\!\!\!\! \!\!\!\!\!\!\!
    \eqfig{0.3 \columnwidth}{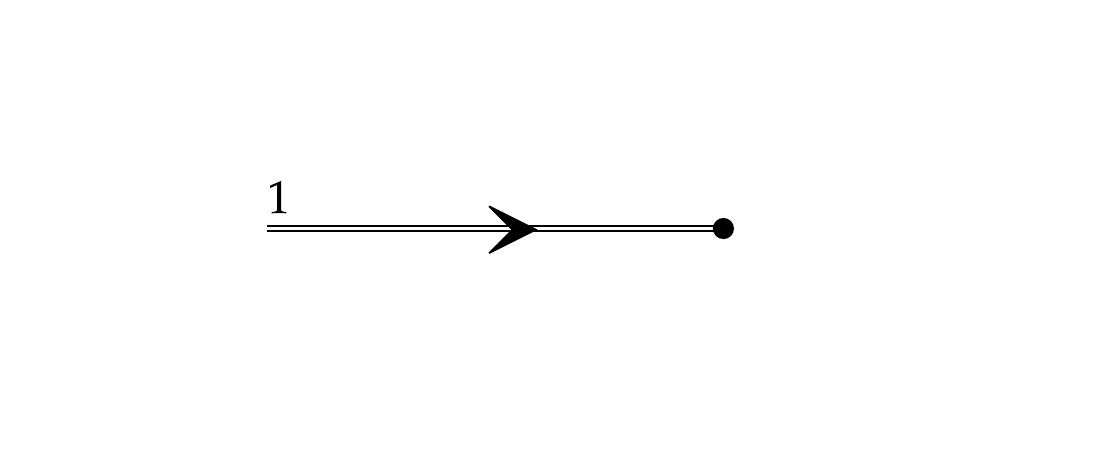}
    \!\!\!\!\!\!\! \!\!\!\!\!\!\!
    &= n_1 - (1{+}n_1) = - 1 
        \label{1v1} 
\end{align}
\vspace{-1cm}
\item Multiply the result by the product of the couplings at  each vertex, and integrate the resulting expression over all internal momenta. 
\item Repeat Steps 1 through 5 for all possible time orderings, sum the results, and include the appropriate Feynman diagram combinatorics factor. 
\end{enumerate}

\subsubsection*{Examples}

Let us see how Step $3$ reproduces (\ref{324}), which was  found for the $s$-channel one loop diagram in Fig.~\ref{Figschan}. For $t_a>t_b$, we begin at the vertex at $t_a$. We have an effective $2$-body vertex, since we only look at the lines going to the external time. From the form in (\ref{2v1}) this gives the factor $-(1{+}n_1{+}n_{2})$. Now, examining  the $b$ vertex, since it is at the earliest time, it is an effective $4$-body vertex. From the form in (\ref{4v}) this gives the factor $n_5 n_6(1{+}n_3)(1{+}n_4) - (1{+}n_5)(1{+} n_6)n_3n_4 $. 

\begin{figure}[h] \centering
 \includegraphics[width=2.2in]{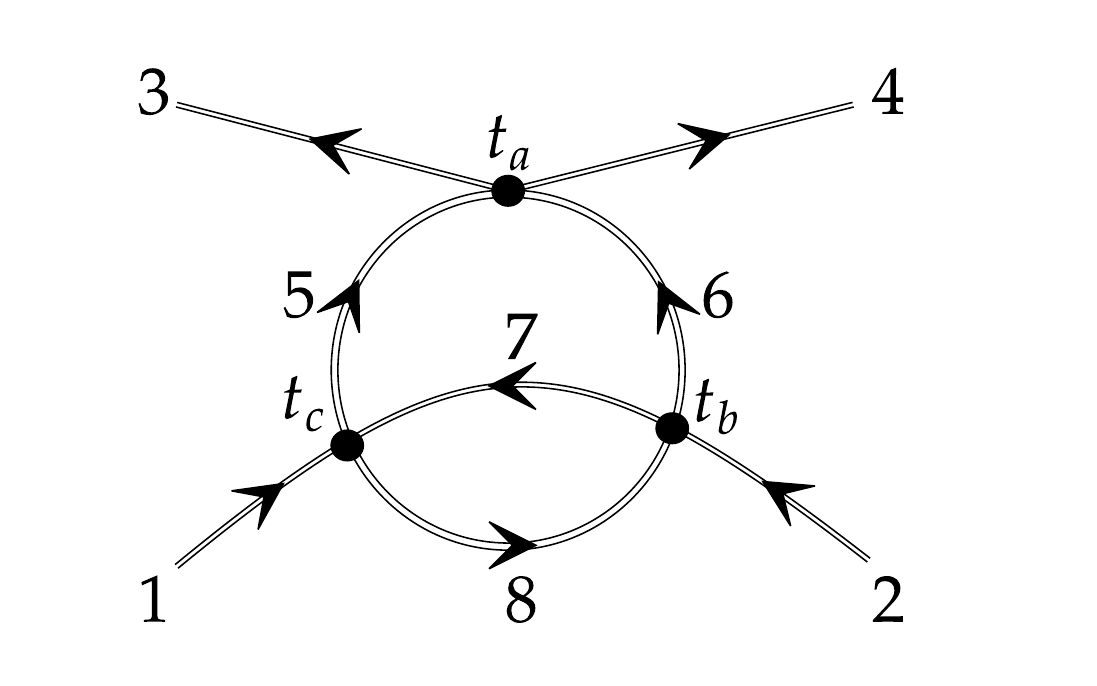} 
\caption{A two-loop diagram.} \label{Fig2loop}
\end{figure}
Let us now give some examples of applying these rules to   higher-loop diagrams. For instance, consider the two-loop diagram  shown in Fig.~\ref{Fig2loop}. The  time ordering $t_a>t_b>t_c$ gives the factor, 
\be
    \frac{1}{\o_{58;17} {-} i\epsilon}  \frac{1}{\o_{56;12} {-} i\epsilon}  \frac{1}{\o_{34;12} {-} i\epsilon}
    (n_3{+}n_4{+}1) (n_2{ -} n_6) \Big(n_1 n_7 (n_5 {+} 1)(n_8 {+} 1) - (n_1 {+} 1)(n_7 {+}1) n_5 n_8\Big)
\ee
On the other hand, the ordering $t_c>t_b>t_a$ gives the factor, 
\be
    -\frac{\Big(n_2 n_8 (n_7 {+}1) - (n_2 {+}1) (n_8 {+}1) n_7 \Big) \Big(n_5 n_6 (n_3 {+}1) (n_4 {+}1) - (n_5{+}1) (n_6{+}1) n_3 n_4\Big)}{(\o_{34;56} {-} i\epsilon) (\o_{347;258}{ -} i\epsilon) (\o_{34;12}{ -} i\epsilon)}~.
\ee
There are four additional time orderings to consider, but we stop here.

\begin{figure}[h] \centering
 \includegraphics[width=2.2in]{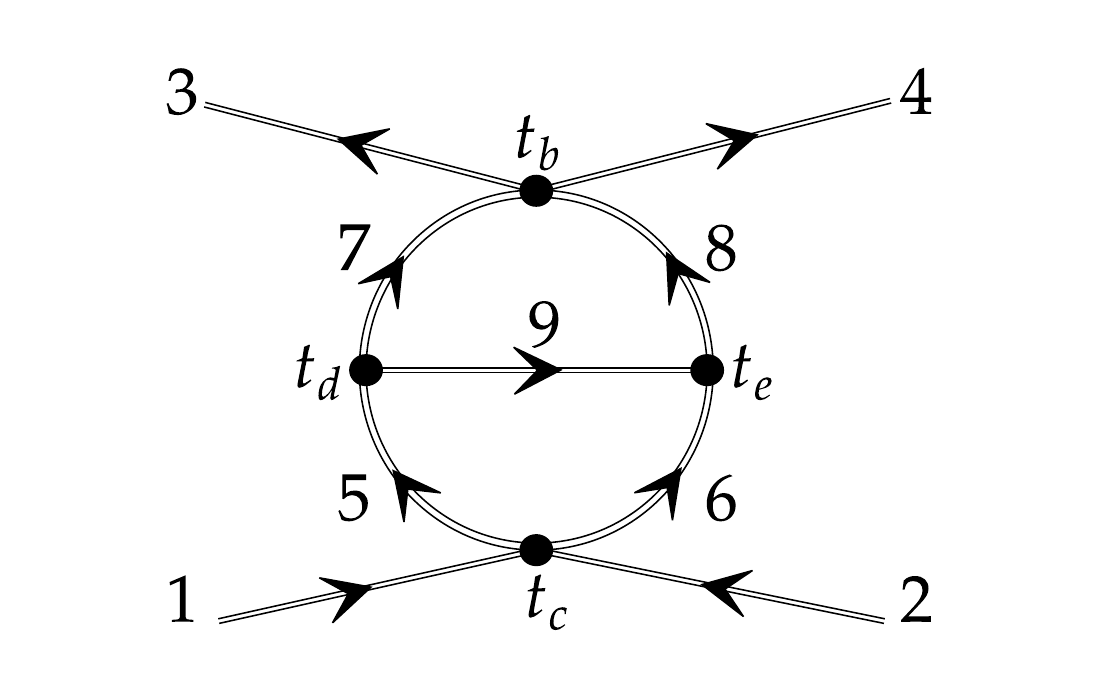} 
\caption{A two-loop diagram in a theory with both cubic and quartic interactions.} \label{Fig2cubic}
\end{figure}

These rules are not specific to quartic interactions. For instance, consider the  diagram in Fig.~\ref{Fig2cubic}, taken from Fig.~6 of \cite{RSSS}, in a theory with both quartic and cubic interactions. An example of a time ordering is $t_b>t_e>t_c>t_d$, which  gives the factor, 
\be
    \frac{(n_3{+}n_4{+}1)\Big(n_1 n_2 (n_6{+}1) - (n_1 {+} 1) (n_2 {+}1) n_6 \Big) \Big( n_5 (n_7{+}1) (n_9{+}1) - (n_5{+}1) n_7 n_9\Big)}{(\o_{79;5}{ -} i\epsilon) (\o_{679;12} {-} i\epsilon) (\o_{78;12} {-} i\epsilon)(\o_{34;12} {-} i\epsilon)}  
\ee

Setting up the calculation of the correlation functions in standard quantum mechanical perturbation theory, as  done in Appendix~\ref{apC},  shows that for each diagram, every  term contains either an $n_i$ or a $(1+n_i)$ factor for each momentum $p_i$ appearing in the diagram. This is just because either the annihilation operator acts first followed by the action of the creation operator, $\da_i a_i |n_i\ra = n_i |n_i\ra$, or vice-versa, $a_i \da_i |n_i\ra = (n_i +1) |n_i\ra$. The power of (\ref{325}) is that it accounts for minus signs, telling us how all these terms combine. 

\section{Discussion} \label{sec4}

The Boltzmann equation describes the rate of change of particle number, accounting for the two-to-two scattering amplitude while neglecting all other multi-particle  scattering processes. As the density increases, it stops being legitimate to ignore such terms. Assuming weak coupling and a state that is nearly  homogenous and stationary, we have explicitly computed the next-order terms in the density, and provided a simple algorithm for writing down terms at any order.

We have worked within the context of quantum field theory, which describes both particles and waves. Classically, the kinetics of waves and particles have qualitatively different structures. The first-principles description of the kinetics of classical particles is through the celebrated BBGKY hierarchy, which relates the rate of change of a one-particle distribution to a two-particle distribution, whose rate of change is, in turn, expressed in terms of a three-particle distribution, and so on. A low density, combined with the assumption of chaos, allows one to truncate the hierarchy by factorizing the two-particle distribution into a product of two one-particle distributions, yielding the Boltzmann equation. The kinetics of classical waves, on the other hand, relates the rate of change of an equal-time two-point correlation function to a four-point correlation function via the equations of motion, whose rate of change is, in turn, expressed in terms of a six-point function, and so forth. Weak interactions,  and a state that is nearly  Gaussian,  allows one to factorize higher-point functions into two-point functions -- occupation numbers. It may be useful to study the classical particle limit of our results for the higher-order kinetic equation for quantum fields. 

The  wave kinetic equation has been of particular interest recently due to the existence of stationary, far-from-equilibrium, scale-invariant  constant-flux solutions, $n_k \sim k^{-\g}$, which appear in a broad range of physical contexts with weakly interacting solutions, such as:  gravity waves in the ocean \cite{Newell, PhysRevLett.134.054101} and capillary waves \cite{FalconMordant, kochurin}, Bose gases \cite{exp2,exp3, Nowak:2011sk, Naz23, hu2024, Rosenhaus:2025mgj, Noel:2025mtb}, and cascades in early universe reheating \cite{Micha:2004bv, Gregory:2024ogi} and following hadron collisions \cite{Berges:2020fwq}. Generally, the classical wave limit ($n_k\gg 1$) is taken, in order to have scale-invariant behavior. Nevertheless, through use of (\ref{kineticEq-formal}), one can search for stationary far-from-equilibrium states in any quantum field theory in which the four-point function can be computed; the prime example so far has been large-$N$ theories \cite{B15, Walz:2017ffj, Gaz19, RSch}.

\sss*{Acknowledgments} 
We thank I.~Rothstein, D.~Schubring, and M.~Smolkin for helpful discussions. XYH is supported by the Shuimu Tsinghua Scholar Program at Tsinghua University and a fellowship from the China Postdoctoral Science Foundation (Certificate No. 2024M761598).  VR is supported by NSF grant 2209116 and by BSF grant 2022113.  XYH thanks the NYU Center for Cosmology and Particle Physics and the CUNY Graduate Center for their hospitality, during which part of this work was carried out. 
\appendix
\section{Contour-ordered propagators}
\label{appendix:green-and-occupationnum}
In this appendix, we review the basics of contour-ordered Green's functions  and their relation to the occupation number of a given mode.

The contour-ordered Green's function $G_{\mathcal{C}}(\tau,\tau') \equiv \big\la T_{\mathcal{C}}\{a(\tau)a^{\dagger}(\tau')\}\big\ra$ is defined as
\be
    \!\!\!\!\!\!\! G_{\mathcal{C}}(\tau,\tau') =\! \begin{cases}
        G^{++}(\tau,\tau') \equiv \big\la T_{\mathcal{C}} \{a^+(\tau) a^{+\dagger}(\tau')\} \big\ra
        = G^T(t,t')
        =\big\la T\{a(t) a^{\dagger}(t')\}\big\ra~, & \tau, \tau'\in \mathcal{C}_+ ~,\\
        G^{+-}(\tau,\tau') \equiv 
        \big\la a^{-\dagger}(\tau')a^{+}(\tau) \big\ra
        =G^<(t,t')
        = \big\la a^{\dagger}(t') a(t)\big\ra ~, & \tau \in \mathcal{C}_+,\ \tau'\in \mathcal{C}_-\\
        G^{-+}(\tau,\tau') \equiv
        \big\la  a^{-}(\tau)a^{+\dagger}(\tau') \big\ra
        =G^>(t,t')
        = \big\la a(t) a^{\dagger}(t') \big\ra ~, & \tau \in \mathcal{C}_-,\ \tau'\in \mathcal{C}_+\\
        G^{--}(\tau,\tau') \equiv \big\la T_{\mathcal{C}} \{a^-(\tau) a^{-\dagger}(\tau')\} \big\ra
        = G^{\bar{T}}(t,t')
        =
        \big\la \bar{T}\{a(t) a^{\dagger}(t')\}\big\ra~, & \tau, \tau'\in \mathcal{C}_-~, 
    \end{cases}
 \label{A1}
\ee
where $a^{\dagger}$ and $a$ are the creation and annihilation operators in the original theory (\ref{Hamiltonian}) and the Schwinger-Keldysh contour is defined as $\mathcal{C} = \mathcal{C}_+ \cup \mathcal{C}_-$, $t$ and $t'$ represent the values of $\tau$ and $\tau'$, respectively, regardless of whether they are on the $\mathcal{C}_+$ or $\mathcal{C}_-$ branch, see Fig.~\ref{FigKeldysh}.
For example, we have the operator relation $a(t) = a^{+}(t) = a^{-}(t)$. The symbol $T$ denotes the time-ordering operator, whereas the $\bar T$ represents the anti-time-ordering opereator.
Here, $\la \mO \ra$ represents the transition amplitude from the initial state $|\Omega\ra$ at $t=t_0$ back to itself, with an operator (in the Heisenberg or interaction picture) inserted somewhere on the contour,  $\la\mO\ra \equiv \la \Omega|\mO|\Omega\ra$.

The contour-ordered Green's function can  naturally be organized into a two-by-two matrix form in the $(a^+,a^-)$ basis,
\begin{align}
    G_{\mathcal{C}} = \begin{pmatrix}
        G^{++} & G^{+-} \\
        G^{-+} & G^{--}
    \end{pmatrix}
    =\begin{pmatrix}
        G^T & G^<\\
        G^> & G^{\bar T}
    \end{pmatrix}~.
\end{align}
By performing the Keldysh rotation (\ref{Keldysh-rotated fields}), we obtain  the matrix in the $(A,\eta)$ basis,
\begin{align}
    G_{\mathcal{C}} = \begin{pmatrix}
        G^K & G^R\\
        G^A & 0
    \end{pmatrix}
\end{align}
where
\bea
    G^K &=& G^> + G^< = G^T + G^{\bar T}~,
    \label{eq:SK-relation-1}
    \\
    G^R &=& G^T - G^< = G^> - G^{\bar T}~,
    \label{eq:SK-relation-2}
    \\
    G^A &=& G^T - G^> = G^< - G^{\bar T}~.
    \label{eq:SK-relation-3}
\eea
The superscripts $K$, $R$, and $A$ denote the Keldysh, retarded, and advanced Green's functions, respectively. By substituting (\ref{A1}) into (\ref{eq:SK-relation-1}) -- (\ref{eq:SK-relation-3}), one gets
\be
  G^K(t,t') = \big\la \{a^{\dagger}(t'), a(t)\}\big\ra~,
  \quad 
  G^R(t,t') = \big\la [a(t),a^{\dagger}(t')]\big\ra \theta(t{-}t')~,
  \quad 
  G^A(t,t') = \big\la[a^{\dagger}(t'),a(t)] \big\ra \theta(t'{-}t)~.
  \label{GKRA_A7}
\ee
The time ordering of these Green's functions is now explicitly defined. It is straightforward to verify that the expressions in (\ref{fig:GK-v2}) -- (\ref{fig:GA-v2}) match those in (\ref{GKRA_A7}).

As indicated by (\ref{A1}), the occupation number $n \equiv \la a^{\dagger}(t)a(t)\ra$ is  given by $G^{<}(t,t')$ in the limit $t'\to t$. 
Using  the relations (\ref{eq:SK-relation-1})--(\ref{eq:SK-relation-3}), we  express the occupation number 
in terms of $G^K$, $G^A$ and $G^R$ as 
\begin{align}
    n = \lim_{t' \to t} G^<(t,t') = \lim_{t' \to t}\frac{1}{2}\left[G^K(t,t') + G^A(t,t') - G^R(t,t')\right]~.
\end{align}

\subsection*{Kinetic equation and equal-time four-point function}
\label{appendix:KE}
Here, we derive the quantum kinetic equation (\ref{kineticEq-formal}), which relates the change in the occupation number to an equal-time four-point function. 

Combining the Heisenberg equation of motion for the Hamiltonian (\ref{Hamiltonian}), 
\be
    \dot{a}_k = i[H,a_k]=
    -i
    \o_k a_k - 2i \int\! \prod_{i=2}^4 d^d k_i\, \lambda_{k234} a^{\dagger}_2 a_3 a_4~,
\ee
 with  its Hermitian conjugate, we obtain the time derivative of the occupation number of mode $k$,  $n_k\equiv \la a_k^{\dagger}(t) a_k(t)\ra$,
\be
    \pdv{n_k}{t} = \la \dot{a}^{\dagger}_k a_k\ra + \la a^{\dagger}_k \dot{a}_k\ra 
    = 2i  \int\! \prod_{i=2}^4 d^d k_i\left(
    \lambda_{k234}^* \la a_4^{\dagger} a_3^{\dagger} a_2 a_k \ra
    - \lambda_{k234} \la a^{\dagger}_k a^{\dagger}_2 a_3 a_4\ra
    \right)~. \label{A10}
\ee
where we used that $[a_i,a_j^{\dagger}] = \delta_{ij}$. 
For an operator $\mO$ and two general states $|\alpha\ra $ and $|\beta\ra$, we have the identity $\la\beta|\mO|\alpha\ra = \la\alpha|\mO^{\dagger}|\beta\ra^*$. 
On the Keldysh contour, $\big\la \mO^{\dagger}\big\ra = \la\mO\ra^*$, and (\ref{A10}) thus reduces to 
\be
    \pdv{n_k}{t} = 4\Im  \int\! \prod_{i=2}^4 d^d k_i\, \lambda_{k234} \big\la a^{\dagger}_k a_2^{\dagger} a_3 a_4\big\ra(t)
    ~,
    \label{kineticEq_B3}
\ee
which corresponds to (\ref{kineticEq-formal}) in the main text. The equal-time four-point function $\la a_1^{\dagger} a_2^{\dagger} a_3 a_4\ra$ can be evaluated using either $a_k^+ = \frac{1}{\sqrt{2}} (A_k + \eta_k)$ or $a_k^- = \frac{1}{\sqrt{2}}(A_k-\eta_k)$; the result should be independent of the choice. Indeed, using four $a^+$ operators as an example,
\be
\la a_1^{\dagger} a_2^{\dagger} a_3 a_4\ra =\la a_1^{+\dagger} a_2^{+\dagger} a^+_3 a^+_4\ra= \frac{1}{4}
    \la A_1^{\dagger} A_2^{\dagger} A_3 A_4\ra~.
\ee
The last equality holds because equal-time correlators involving any number of $\eta$ fields, such as e.g. $\la A_1^{\dagger} \eta_2^{\dagger} A_3 A_4\ra$, vanish due to conflicting time orderings imposed by the retarded and advanced Green’s functions,  $G^A(t,t') G^R(t',t) = 0$. This demonstrates the last equality in (\ref{kineticEq-formal}). 

\section{Perturbative computation of equal-time correlation functions} \label{apC}
In this appendix, we compute the equal-time four-point function entering the kinetic equation directly using perturbation theory, to second order. This is  essentially the same derivation as in the main body of the text, but it doesn't make use of the path integral or the Keldysh contour. This provides a complementary perspective on the derivation and the result. Of course, beyond one loop the path integral approach in the main body is more efficient.~\footnote{A similar, but perhaps more involved, approach to the one in this appendix -- perturbatively solving the Heisenberg equation for the density matrix -- was discussed in \cite{Snoke:2012zz}. For an equal-time approach, more similar to the one in this paper, see  \cite{Ott:2022tqd}.}

We will work in the interaction picture: the operators evolve with the free Hamiltonian, 
\be \label{C1}
a_{k}(t) = e^{i H_0 (t-t_0)} a_k e^{-i H_0 (t-t_0)}~,
\ee
while the state evolves with the interaction Hamiltonian, and at time $t$ is given by
\be
|\Psi(t)\ra  = U(t,t_0) | \Psi(t_0)\ra~,
\ee
where the initial state $| \Psi(t_0)\ra $ is one in which mode $k$ has occupation number $n_k$, i.e., it is a product of harmonic oscillator states $|n_k\ra$ for each mode $k$. The evolution operator is given by,
\be
 U(t,t_0) = T \exp\( - i \int_{t_0}^t dt' H_{int}(t’)\)  = 1 - i \int_{t_0}^t d t_a H_{int}(t_a) - \int_{t_0}^t d t_a \int_{t_0}^{t_a} d t_b \, H_{int}(t_a) H_{int}(t_b) + \ldots
 \ee
where $T$ denotes time ordering, meaning that operators at later times appear further to the left.

We would like to compute the equal-time four-point function $\la a_1^{\dagger} a_2^{\dagger} a_3 a_4\ra$, where the operator is at time $t$. 
Denoting 
\be
V_{1234} (t) \equiv a_1^{\dagger}(t) a_2^{\dagger}(t) a_3(t) a_4(t)
\ee
so that $H_{int} =\sum_{1,2,3,4} \lam_{1234} V_{1234}$
we need to compute, 
\be
\la a_1^{\dagger} a_2^{\dagger} a_3 a_4\ra\equiv \la \Psi(t) | V_{1234} (t) |\Psi(t)\ra = \la \Psi(t_0)| U(t,t_0)^{\dagger} V_{1234}(t) U(t,t_0)|\Psi(t_0)\ra
\ee
This is, of course, the Keldysh contour procedure used in the main text: the $U(t,t_0)$ is the portion of the contour running forwards in time, while $U(t,t_0)^{\dagger} $ runs backwards in time. Here, we will perform the computation explicitly to the first two orders, without using Feynman rules or propagators.

To order $\lam$ we get, 
\bea \nn
\!\!&&\la a_1^{\dagger} a_2^{\dagger} a_3 a_4\ra = - i\int_{t_0}^t dt'  \( \la \Psi(t_0)| V_{1234}(t) H_{int}(t')|\Psi(t_0)\ra - \la \Psi(t_0)|  H_{int}(t') V_{1234}(t)|\Psi(t_0)\ra\)+\ldots \\ \label{C6}
&=&- 4i\lam_{3412}\int_{t_0}^t dt' \! \( \la \Psi(t_0)| V_{1234}(t) V_{3412}(t')|\Psi(t_0)\ra - \la \Psi(t_0)|  V_{3412}(t') V_{1234}(t)|\Psi(t_0)\ra\)+\ldots
\eea
We note that the action of an annihilation operator for a single mode  (\ref{C1}) on the initial state is
\be
a_k(t) |n_k\ra = e^{i \o_k (n_k -1)(t -t_0)} \sqrt{n_k} e^{-i \o_k n_k (t-t_0)} |n_k {-}1\ra = e^{-i \o_k (t-t_0)} \sqrt{n_k}|n_k{-}1\ra
\ee
since the energy of the free system before $a_k$ acts is $\o_k n_k$, whereas after $a_k$ acts,  it is $\o_k (n_k{-}1)$. Thus, we obtain, 
\bea
\int_{t_0}^t dt' \la \Psi(t_0)|V_{1234}(t) V_{3412}(t')| \Psi(t_0)\ra \nn
 &=&n_1 n_2 (n_3{+}1)(n_4{+}1)e^{i \o_{12;34}(t-t_0)} \int_{t_0}^t dt' e^{ i \o_{34;12}(t'-t_0)}\\ 
 & =&  -i\frac{n_1 n_2 (n_3{+}1)(n_4{+}1)}{\o_{34;12}{-}i\eps}
\eea
Therefore, (\ref{C6})  reproduces the tree-level answer (\ref{34}). 

At second order, there are three terms: 
\begin{enumerate}
\item  
\be \label{C9}
-\int_{t_0}^t d t_a \int_{t_0}^{t_a} d t_b\,  \la \Psi(t_0)| V_{1234}(t)  H_{int}(t_a) H_{int}(t_b)|\Psi(t_0)\ra
\ee
\item  
\be \label{C10}
-\int_{t_0}^t d t_a \int_{t_0}^{t_a} d t_b\,  \la \Psi(t_0)| H_{int}(t_b) H_{int}(t_a)  V_{1234}(t) |\Psi(t_0)\ra
\ee
\item 
\be \label{C11}
\int_{t_0}^t d t_a \int_{t_0}^{t} d t_b\,  \la \Psi(t_0)| H_{int}(t_a) V_{1234}(t)  H_{int}(t_b)|\Psi(t_0)\ra
\ee
\end{enumerate}
We will first focus on obtaining the $s$-channel four-point function. 
Starting with the first term, we see that the two $H_{int}$ must be $V_{5612}$ and $V_{3456}$, with the only choice being which one is at time $t_a$ and which is at $t_b$. So (\ref{C9}) becomes, 
\small
\bml \label{C12}
\!\!\!\!\!\!\!\!\!-16 \lam_{3456} \lam_{5612}\!\int_{t_0}^t \!d t_a \!\!\int_{t_0}^{t_a}\! \!d t_b \Big(\! \la \Psi(t_0)| V_{1234}(t) V_{5612}(t_a) V_{3456}(t_b)|\Psi(t_0)\ra + \la \Psi(t_0)| V_{1234}(t) V_{3456}(t_a) V_{5612}(t_b)|\Psi(t_0)\ra \!\Big) \\
=  16\lam_{3456} \lam_{5612}\frac{n_1 n_2 (n_3{+}1)(n_4{+}1) }{\o_{34;12}{-}i\eps} \(\frac{ n_5 n_6}{\o_{34;56}{-}i\eps}+\frac{(n_5{+}1) (n_6{+}1)}{\o_{56;12}{-}i\eps}\)
\end{multline}
\normalsize
Likewise, (\ref{C10}) becomes
\small
\bml \label{C13}
\!\!\!\!\!\!\!\!-16 \lam_{3456} \lam_{5612}\!\int_{t_0}^t\! \!d t_a\! \int_{t_0}^{t_a}\!\! d t_b \Big(\! \la \Psi(t_0)|  V_{5612}(t_b) V_{3456}(t_a)V_{1234}(t)|\Psi(t_0)\ra + \la \Psi(t_0)|V_{3456}(t_b) V_{5612}(t_a)  V_{1234}(t) |\Psi(t_0)\ra\! \Big) \\
=  16\lam_{3456} \lam_{5612}\frac{(n_1{+}1) (n_2{+}1)n_3n_4 }{\o_{34;12}{-}i\eps} \(\frac{ n_5 n_6}{\o_{56;12}{-}i\eps}+\frac{(n_5{+}1) (n_6{+}1)}{\o_{34;56}{-}i\eps}\)
\end{multline}
\normalsize
Finally, for (\ref{C11}), there is no time ordering and we get, 
\small
\bea\nn
&&\!\!\!\!\!\!\!\!\!\!\!\!\!\!\!16 \lam_{3456} \lam_{5612}\int_{t_0}^t\! d t_a\! \int_{t_0}^{t}\! d t_b\Big(\! \la \Psi(t_0)|  V_{5612}(t_a) V_{1234}(t) V_{3456}(t_b)|\Psi(t_0)\ra{ +} \la \Psi(t_0)| V_{3456}(t_a) V_{1234}(t) V_{5612}(t_b)|\Psi(t_0)\ra \!\Big) \\  \label{C14}
&=& \!\! -16 \lam_{3456} \lam_{5612}\frac{(n_1{+}1)(n_2{+}1)(n_3{+}1)(n_4{+}1)n_5 n_6 + n_1n_2 n_3 n_4 (n_5{+}1)( n_6{+}1)  }{(\o_{34;56}{-}i\eps)(\o_{56;12}{-}i\eps)}\\ \nn
&=&\!  \! -16\lam_{3456} \lam_{5612}\frac{(n_1{+}1)(n_2{+}1)(n_3{+}1)(n_4{+}1)n_5 n_6 + n_1n_2 n_3 n_4 (n_5{+}1)( n_6{+}1)  }{\o_{34;12}{-}i\eps}\Big( \frac{1}{\o_{34;56}{-}i\eps} + \frac{1}{\o_{56;12}{-}i\eps}\Big) \nn
\eea
\normalsize
where in the last equality we rewrote the $\o$ denominators in a way that will be convenient in what we do next. 

Let us combine the terms from (\ref{C12}),( \ref{C13}), (\ref{C14}) that have $\o_{34;56}$ in the denominator. This gives, 
\bml
\frac{ 16 \lam_{3456} \lam_{5612} }{(\o_{34;12}{-}i\eps)(\o_{34;56}{-}i\eps)}\Big( n_1 n_2 (n_3{+}1)(n_4{+}1) n_5 n_6+ (n_1{+}1) (n_2{+}1)n_3n_4 (n_5{+}1) (n_6{+}1)\\-(n_1{+}1)(n_2{+}1)(n_3{+}1)(n_4{+}1)n_5 n_6 - n_1n_2 n_3 n_4 (n_5{+}1)( n_6{+}1)\Big)
\end{multline}
Combining the numerators factors precisely reproduces what we had in (\ref{s-channel-total}). The same applies to the terms with  $\o_{56;12}$ in the denominator. We thus recover the $s$-channel contribution to the four-point function at one loop, given in  (\ref{s-channel-total}). The $t$-channel contribution is obtained from (\ref{C9}--\ref{C11}) by letting one of the $H_{int}$ have a $V_{3516}$ and the other a $V_{4625}$.

\section{Particle number non-conserving interactions} \label{apB}
While the main body of the text dealt with quartic interactions involving two creation and two annihilation operators (\ref{Hamiltonian}), this can be easily generalized to cubic or higher-order interactions, as well as interactions with an unequal number of creation and annihilation operators.

\subsubsection*{Cubic Interaction}
For instance, consider a  cubic interaction, 
\be
H_{int} = 
\frac{1}{2}\int \prod_{i=1}^3 d^d k_i\,  \lam_{123}\, \da_{k_1} a_{k_2} a_{k_3}  \delta(\v k_{1;23})+ \text{h.c.}~.
\ee
From (\ref{210}), we have that on the Keldysh contour, $L_{int} = - H_{int}(a^+) + H_{int}(a^-)$. Rotating to the $A$ and $\eta$ fields (\ref{Keldysh-rotated fields}) gives,
\be \label{C2}
L_{int} = -\frac{1}{2\sqrt{2}}\int \prod_{i=1}^3d^d k_i\,  \lam_{123} \( \eta_1^{\dagger} \eta_2 \eta_3 + \eta_1^{\dagger}A_2 A_3 + 2A_1^{\dagger} A_2 \eta_3\) +\text{h.c.}~.
\ee
Computing the tree-level,  equal-time three-point function,
\bea \nn
\la A_1^{\dagger} A_2 A_3 \ra_{\text{tree}} &=&  \frac{1}{\sqrt{2}} \lam_{123}^*\big( 1+(2n_2{+}1)(2n_3{+}1)- (2n_1{+}1)(2n_2{+}1)+(2n_1{+}1)(2n_3{+}1) \big) \frac{1}{\o_{23;1}{-} i\eps}\\ \label{B3}
 &=& 2 \sqrt{2} \lam_{123}^*\big( n_2 n_3 (1{+}n_1) - n_1(1{+}n_2) (1{+}n_3) \big) \frac{1}{\o_{23;1}{-} i\eps}~,
\eea
where in the first equality, the first term in parentheses (the $1$) originates from the Hermitian conjugate of the first term ($\eta^3$) in (\ref{C2}), while the second and third terms arise from the Hermitian conjugate of the second and third terms in (\ref{C2}), respectively. 
The kinetic equation is expressed in terms of the correlator as, 
\be 
\frac{\d n_k}{\d t} =\Im  \int \prod_{i=1}^3 d^d k_i \, \(\delta_{k k_1} {-}\delta_{k k_2} {-} \delta_{k k_3}\)\lam_{123} \la \da_{k_1} a_{k_2} a_{k_3}   \ra\delta(\v k_{1;23})~,
\ee
where $\delta_{k k_i} \equiv \delta(\v k{ -}\v k_i)$. Using the three-point function (\ref{B3}), this  gives
\be
\frac{\d n_k}{\d t} = \pi\int \prod_{i=1}^3 d^d k_i  |\lam_{123}|^2 \(\delta_{k k_1} {-}\delta_{k k_2} {-} \delta_{k k_3}\)\delta(\o_{1;23})\delta(\v k_{1;23})\Big(  (1{+}n_1)n_2 n_3 - n_1(1{+}n_2) (1{+}n_3) \Big)
\ee

\subsubsection*{Quartic interaction}
Now consider a quartic interaction, but one that has three creation operators and one annihilation operator, 
\be
H_{int} = 
\frac{1}{2}\int \prod_{i=1}^4d^d k_i\,  \lam_{1;234} \da_{k_1} a_{k_2} a_{k_3} a_{k_4}  \delta(\v k_{1;234})+ \text{h.c.}~.
\ee
From (\ref{210}), we have that on the Keldysh contour, $L_{int} = - H(a^+) + H(a^-)$. Rotating to the $A$ and $\eta$ fields (\ref{Keldysh-rotated fields}) gives
\be
L_{int} = -\frac{1}{4}\int \prod_{i=1}^4d^d k_i\,   \lam_{1;234} \Big( \eta_1^{\dagger} A_2 A_3 A_4{+} 3A_1^{\dagger} A_2 \eta_3 A_4
+ A_1^{\dagger} \eta_2 \eta_3 \eta_4 {+} 3 \eta_1^{\dagger} \eta_2 A_3 \eta_4 \Big) +\text{h.c.}~. 
\ee
The correlator is then, 
\be
\la A_1^{\dagger} A_2 A_3 A_4 \ra_{\text{tree}}= 12\lam_{1;234}^*\Big( (1{+}n_1)n_2 n_3 n_4 - n_1 (1{+}n_2)(1{+}n_3)(1{+}n_4)\Big)\frac{1}{\o_{234;1}{-}i\eps}~,
 \ee
 which, upon inserting into, 
 \be 
\frac{\d n_k}{\d t} =\Im  \int \prod_{i=1}^4 d^d k_i \,\(\delta_{k k_1} {-}3\delta_{k k_2}\)  \lam_{1;234} \la \da_{k_1} a_{k_2} a_{k_3} a_{k_4}\ra
\ee
 gives the tree-level kinetic equation,
\be
\frac{\d n_k}{\d t} =  3\pi\int \prod_{i=1}^4 d^d k_i   |\lam_{1;234}|^2\(\delta_{k k_1} {-}3\delta_{k k_2}\)\delta(\o_{1;234})\delta(\v k_{1;234})\Big(  (1{+}n_1)n_2 n_3 n_4- n_1(1{+}n_2) (1{+}n_3) (1{+}n_4)\Big)
\ee
In fact, we could have  obtained this more simply from the kinetic equation with the particle number conserving quartic interaction (\ref{tree-level-KE}), 
\bml
\frac{\d n_k}{\d t} =  4\pi\!\!\int \prod_{i=1}^4 d^d k_i   |\lam_{1234}|^2\(\delta_{k k_1} {+}\delta_{k k_2}{-}\delta_{k k_3}{-}\delta_{k k_4}\)\Big(  (1{+}n_1)(1{+}n_2 )n_3 n_4- n_1n_2 (1{+}n_3) (1{+}n_4)\Big)\\
\delta(\o_{12;34})\delta(\v k_{12;34})
\end{multline}
by  sending $2\rightarrow -2$, which corresponds to $n_2\rightarrow -1 {-} n_2$ (and accounting for an overall minus sign, and the change in the prefactor).

\subsection{Next-to-leading order kinetic equation for $\lam \phi^4$ theory} \label{sec:phi4}
\begin{figure}[]
\centering
\subfloat[]{
\includegraphics[width=1.4in]{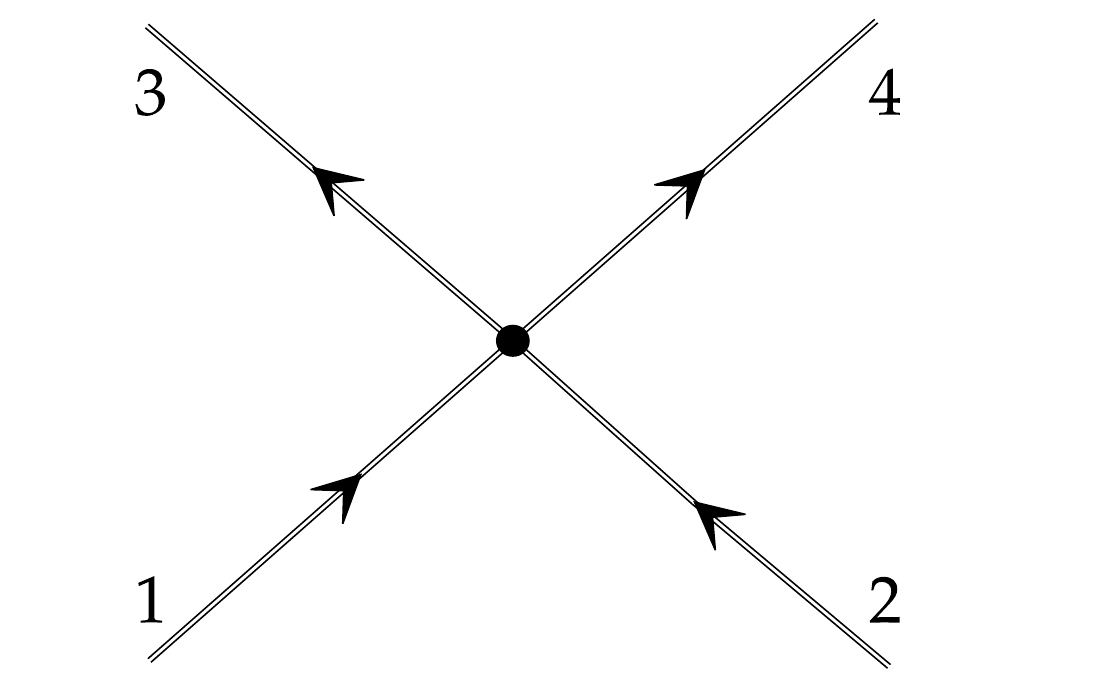}
} 
\subfloat[]{ \includegraphics[width=1.6in]{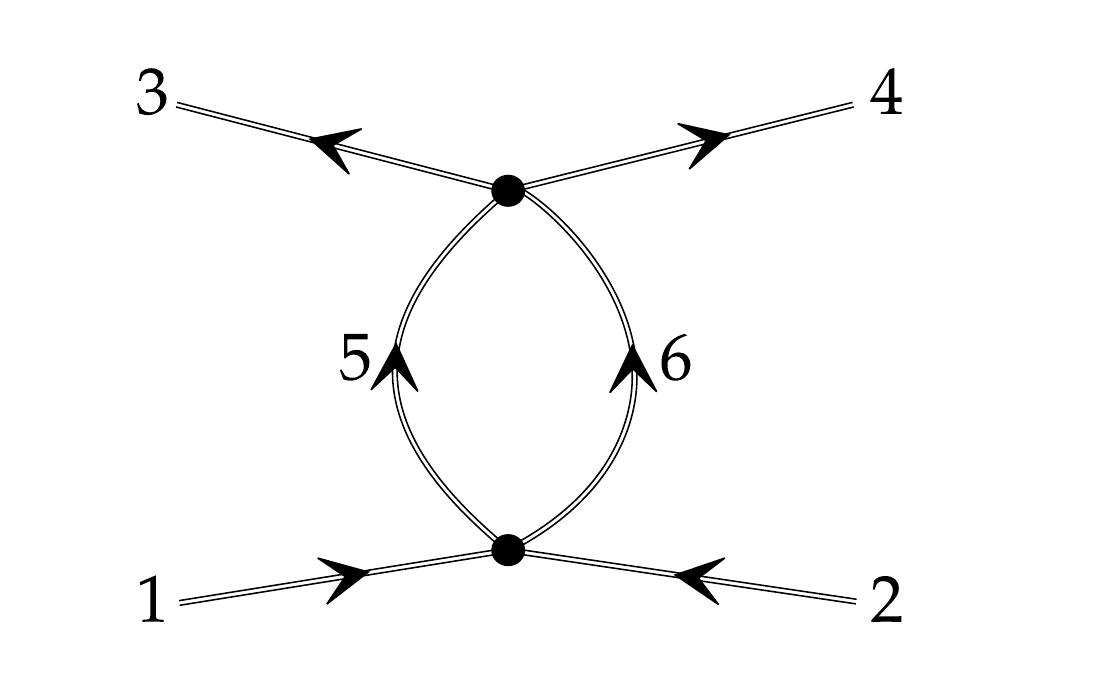}}
   \subfloat[]{ \includegraphics[width=1.6in]{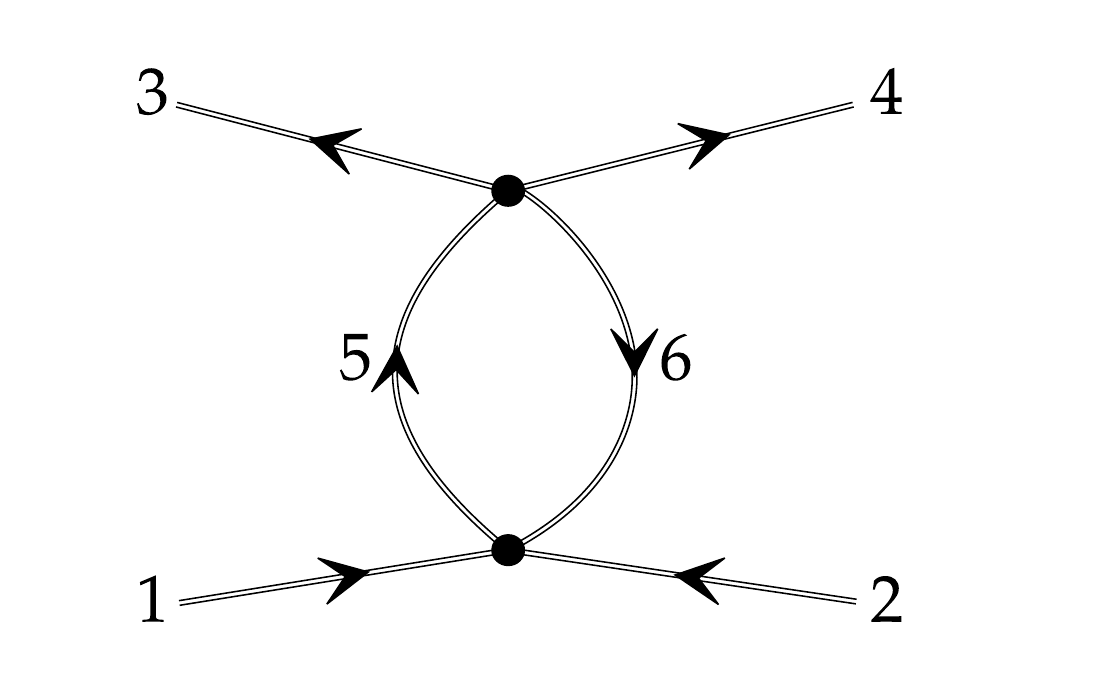}}   \subfloat[]{\ \includegraphics[width=1.6in]{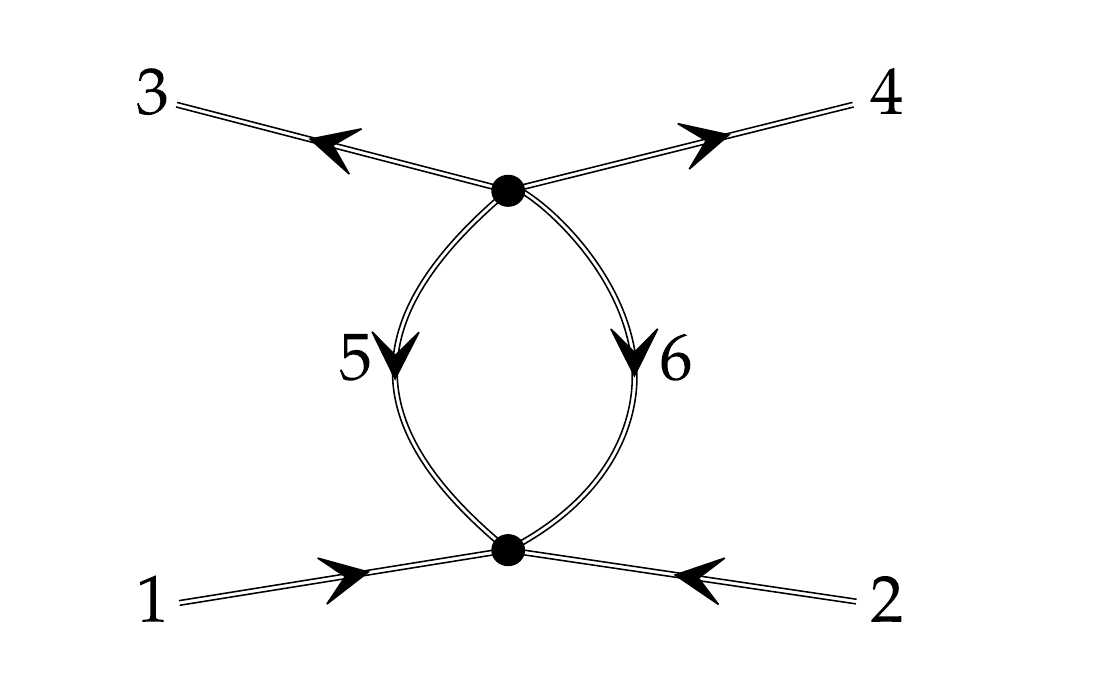}}\\
\subfloat[]{ \includegraphics[width=1.6in]{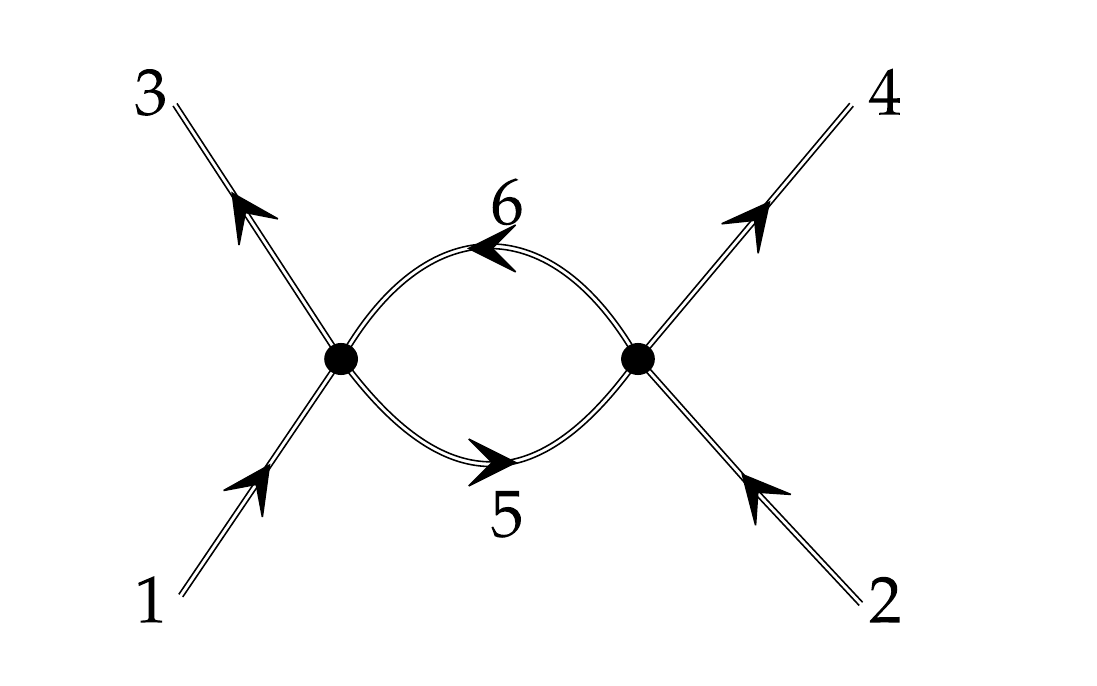}}
 \subfloat[]{ \includegraphics[width=1.6in]{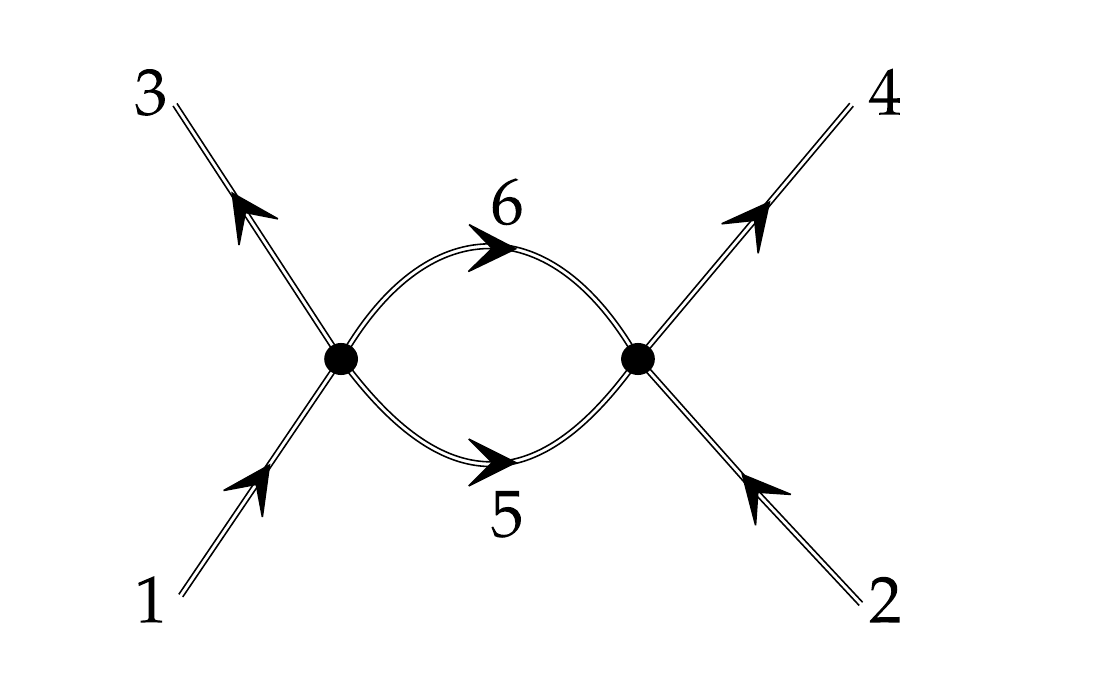}} 
  \subfloat[]{ \includegraphics[width=1.6in]{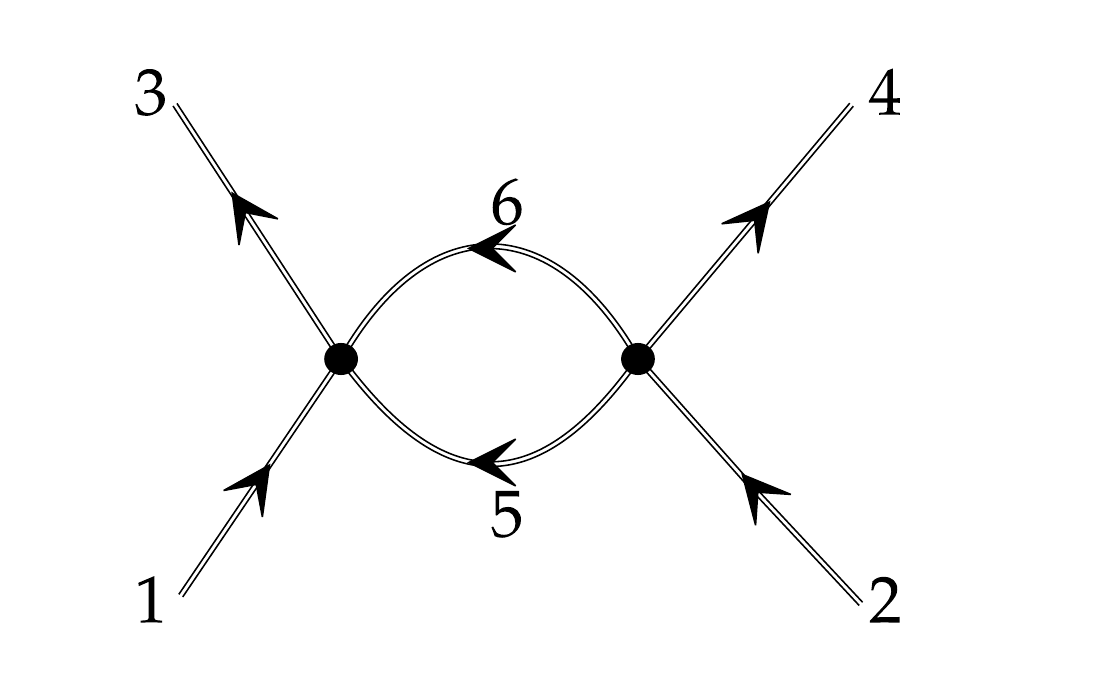}} 
  \caption{Diagrams for the $\la \da_1 \da_2 a_3 a_4\ra$ four-point function in $\lam \phi^4$ theory. (a) tree-level.  One-loop: (b) $\mL_+$, (c) $\bar{\mL}_+$, (d) $\tilde{\mL}_+$, (e) $\mL_-$, (f)$\bar{\mL}_-$, (g) $\tilde{\mL}_-$~. } \label{phi4fig1}
\end{figure}

The bulk of the paper focused on interactions of the form (\ref{Hamiltonian}), which have a definite number of creation and annihilation operators. We may easily extend this to interactions that are sums of such terms. Here we consider the case of relativistic $\lam \phi^4$ field theory, having the interaction (\ref{23}). 
This interaction contains, in addition to $2$ to $2$ scattering, scattering that is $3$ to $1$ or $4$ to $0$. At tree level these other terms are irrelevant, since $3$ to $1$ scattering is not possible for a relativistic scalar with the dispersion relation $\o_k = \sqrt{{\v k}^2 +m^2}$. The only diagram is the one shown in Fig.~\ref{phi4fig1}(a).  At one-loop level, on the other hand, these other interactions can appear and are shown in Fig.~\ref{phi4fig1}. 

As one sees from (\ref{21}) and (\ref{23}), the interaction $\lam_{1234}$ for $\lam \phi^4$ theory is, 
\be
\lam_{1234} = \frac{\lam}{16\sqrt{\o_{k_1} \o_{k_2} \o_{k_3} \o_{k_4}}}
\ee
We find that the next-to-leading-order kinetic equation is,
 \bml \label{B14}
\o_1\frac{\d n_{1}}{dt} = \frac{\pi \lam^2}{16} \int \frac{d^d k_2}{\o_2} \frac{d^d k_3}{\o_3} \frac{d^d k_4}{\o_4}
\Big((n_1{+}1)(n_{2}+1)n_3 n_4-n_1 n_2(n_3{+}1)(n_4{+}1)  \Big) \\
\Big(1 + 2(\mL_+ +2\bar{\mL}_++\tilde{\mL}_+)+ 4(2\mL_-+\bar{\mL}_- +\tilde \mL_-)\Big) \delta(\o_{12;34})\delta(\v k_{12;34}) 
\Big]~.
\end{multline}
Here, the additional loop integrals account for some of the interactions being $3$ to $1$ or $4$ to $0$. Namely, 
the loop integrals $\mL_{\pm}$ which we had before (and correspond to $2$ to $2$ scattering), see (\ref{Lpmin}) are, 
\be
\mL_+ = 2\lam\int\!\! \frac{d^d k_5}{2\o_5} \frac{d^d k_6}{2\o_6}\frac{n_5{+}n_6{+}1}{\o_{12;56} }\delta(\v k_{12;56})~, \ \ \ \ \mL_-= 2\lam\int\!\! \frac{d^d k_5}{2\o_5} \frac{d^d k_6}{2\o_6}\frac{n_6{-}n_5}{\o_{16;35} }\delta(\v k_{16;35})~.
\ee
The diagram $\mL_+^1$ corresponds to replacing the $2$ to $2$ interaction in $\mL_+$ with a $3$ to $1$ interaction, which we achieve by flipping the arrows on $6$, while $\mL_+^0$ corresponds to having a $4$ to $0$ interaction, which we achieve by flipping  both the $5$ and $6$ arrows on $\mL_+^2$, 
\bea 
\bar{\mL}_+ = 2\lam\int\!\! \frac{d^d k_5}{2\o_5} \frac{d^d k_6}{2\o_6}\frac{n_6{-}n_5}{\o_{126;5}}\delta(\v k_{126;5})~,  \ \ \ \ \ \ \ \tilde{\mL}_+ = -2\lam\int\!\! \frac{d^d k_5}{2\o_5} \frac{d^d k_6}{2\o_6}\frac{n_5{+}n_6{+}1}{\o_{1256} }\delta(\v k_{1256})~,
 \eea
 where $\o_{126;5} \equiv \o_1{+}\o_2{+}\o_6{-}\o_5$ and  $\o_{1256} \equiv \o_1{+}\o_2{+}\o_5{+}\o_6$.
We may similarly take  $\mL_-$ and  flip the arrows on $6$, giving $\bar{\mL}_-$, or instead on $5$, giving $\tilde{ \mL}_-$. Both of these have  $3$ to $1$ interactions, 
 \bea
  \bar{\mL}_-= 2\lam\int\!\! \frac{d^d k_5}{2\o_5} \frac{d^d k_6}{2\o_6}\frac{1{+}n_5{+}n_6}{\o_{1;356}}\delta(\v k_{1;356})~
 , \ \ \ \  \tilde{ \mL}_-= -2\lam\int\!\! \frac{d^d k_5}{2\o_5} \frac{d^d k_6}{2\o_6}\frac{1{+}n_5{+}n_6}{\o_{156;3} }\delta(\v k_{156;3})~.
\eea

\subsubsection*{Manifest Lorentz invariance}
Let us rewrite (\ref{B14}) in a way that makes the underlying Lorentz invariance of the theory manifest. 
We  perform variable changes to rewrite the loop integrals as, 
\be \nn
\mL_+ = \lam\int\!\! \frac{d^d k_5}{\o_5 \o_6}\frac{n_5{+}\frac{1}{2}}{\o_{12;56}}~, \ \ \ \ \bar{\mL}_+ = \frac{\lam}{2}\!\int\!\! \frac{d^d k_5(n_5{+}\frac{1}{2})}{\o_5 \o_6}\Big(\frac{1}{\o_{125;6}}{-}\frac{1}{\o_{126;5}}\Big)~,  \ \ \ \tilde{\mL}_+ = -\lam\int\!\! \frac{d^d k_5}{\o_5 \o_6}\frac{n_5{+}\frac{1}{2}}{\o_{1256}}~,
\ee
where $\v k_6 = \v k_1{+}\v k_2{-}\v k_5$ for all the terms. As a result,
\be
\mL_+ +2\bar{\mL}_++\tilde{\mL}_+=2 \lam \int\!\! \frac{d^d k_5(n_5{+}\frac{1}{2})}{\o_5}\(\frac{1}{\o_{12;5}^2{-}\o_6^2} {+}\frac{1}{\o_{125}^2 {-}\o_6^2}\)~.
\ee
Likewise, 
\be \nn
\mL_- = \frac{\lam}{2}\int\!\! \frac{d^d k_5(n_5{+}\frac{1}{2})}{\o_5 \o_6}\Big(\frac{1}{\o_{15;36}}{-}\frac{1}{\o_{16;35}}\Big)~, \ \ \ \bar{\mL}_- = \lam\int\!\! \frac{d^d k_5}{\o_5 \o_6}\frac{n_5{+}\frac{1}{2}}{\o_{1;356}}~, \ \ \ \tilde{\mL}_- = {-}\lam\!\!\int\!\! \frac{d^d k_5}{\o_5 \o_6}\frac{n_5{+}\frac{1}{2}}{\o_{156;3}}~,
\ee
where $\v k_6 =  \v k_1{-}\v k_3{-}\v k_5$ for all the terms. As a result,
\be
2\mL_- +\bar{\mL}_-+\tilde{\mL}_- =2 \lam \int\!\! \frac{d^d k_5(n_5{+}\frac{1}{2})}{\o_5}\(\frac{1}{\o_{1;35}^2{-}\o_6^2} {+}\frac{1}{\o_{15;3}^2 {-}\o_6^2}\)~.
\ee
Therefore, we can rewrite (\ref{B14}) as \cite{Walz:2017ffj}
 \bml \label{B22}
\o_{|\bf k_1|}\frac{\d n_{1}}{dt} = \frac{\pi \lam^2}{16} \int \frac{d{ \bf k_2}}{\o_{|\bf k_2|}} \frac{d {\bf k_3}}{\o_{|\bf k_3|}} \frac{d{\bf k_4}}{\o_{|\bf k_4|}}
\Big((n_1{+}1)(n_{2}+1)n_3 n_4-n_1 n_2(n_3{+}1)(n_4{+}1)  \Big) \\
\Big(1 + 2\mL(k_1{+}k_2)+4\mL(k_1{-}k_3)\Big) \delta(\o_{|\bf k_1|}{+}\o_{|\bf k_2|}{-}\o_{|\bf k_3|}{-}\o_{|\bf k_4|})\delta({\bf k_1}{+}{\bf k_2}{-}{\bf k_3}{-}{\bf k_4}) 
\Big]~,
\end{multline}
where $k$ is a four-vector, with time component $k^0$ and spatial component ${\bf k}$, and 
\be
\mL(k) = \lam\!\!\int  d^d q\frac{(n_{\bf q}{+}\frac{1}{2}) }{2 \o_{|{\bf q}|}}\Big[\frac{1}{(k^0 {-}\o_{ |{\bf q}|})^2{ -} ({\bf k} {-}{\bf q})^2 }+\frac{1}{(k^0 {+} \o_{|{\bf q}|})^2 {-} ({\bf k} {-}{\bf q})^2 }\Big]  
~,
\ee
which we may rewrite in terms of a $d{+}1$ spacetime dimension integral, 
\be \label{mLp}
\mL(k) = 2 \lam\!\!\int\!\! d^{d+1} q \frac{(n_{\bf q}{+}\frac{1}{2})\delta(q^2 {-} m^2) }{(k{-}q)^2 }~,
\ee
where we are using relativistic notation, $k^2 = k_0^2 - {\bf k}^2$.

The expression (\ref{B22}) naturally emerges from working with $\phi$ fields rather than with $a$ fields. Namely, upon doing a field rotation, see e.g. \cite{Mueller:2002gd,Jeon:2004dh},
\be
\phi = \frac{1}{2}(\phi^+ + \phi^-)~, \ \ \ \ \ \eta = \phi^+- \phi^-~,
\ee
the $\lam \phi^4$ Lagrangian on the Keldysh contour takes the form,
\be \label{etaphiL}
\mL =\d \eta \d \phi - \frac{\lam}{3!} (\eta \phi^3+  \frac{1}{4}\eta^3 \phi)~,
\ee
and the Keldysh and retarded Green's functions take the form, respectively, 
\be\label{GkR}
G_k^K= \la \phi_k^* \phi_k\ra = \delta(k^2{-}m^2)(n_{\bf k}+\frac{1}{2})~, \ \ \ G^R_k = \la \phi_k^* \eta_k\ra = \frac{i}{k^2{-}m^2 {+} i \eps k_0}~.
\ee
 The Feynman diagrams now no longer have arrows and the one-loop contribution to  the $\eta \phi^3$ vertex is given by 
\be
\lam \mL(k_1{+}k_2) + \lam\mL(k_1{-}k_3) + \lam\mL(k_1{-}k_4)
 \ee
 where $\mL$ is given by (\ref{mLp}) and contains a Keldysh propagator for one of the internal lines and a retarded propagator for the other (\ref{GkR}). The $u$ channel contribution is equivalent to the  $t$ channel contribution after a $3\leftrightarrow 4$ variable change. Thus we get the one-loop kinetic equation (\ref{B22}).

\bibliographystyle{utphys}

\end{document}